\documentclass[a4,referee,pre,showpacs]{revtex4}

\usepackage[dvips]{graphicx}
\usepackage{amsmath}

\def\EQ{\begin{equation}}
\def\EN{\end{equation}}
\def\EQA{\begin{eqnarray}}
\def\ENA{\end{eqnarray}}
\def\uu{{\bf u}}
\def\vv{{\bf v}}

\def\UU{{\bf U}}

\def\A{\mathcal{A}}

\def\OOmega{{\mbox{\boldmath$\Omega$}}}
\def\OB{\bar{\Omega}}
\def\OBN{\Omega_0}

\def\T{\mathcal{T}}

\begin{document}
\title{Analytical theory of forced rotating sheared turbulence. I. Perpendicular case}

\author{Nicolas Leprovost and Eun-jin Kim}
\affiliation{Department of Applied Mathematics, University of Sheffield, Sheffield S3 7RH, UK}

\begin{abstract}
Rotation and shear flows are ubiquitous features of many astrophysical and geophysical bodies. To understand their origin and effect on turbulent transport in these systems, we consider a forced turbulence and investigate the combined effect of rotation and shear flow on the turbulence properties. Specifically, we study how rotation and flow shear influence the generation of shear flow (e.g. the direction of energy cascade), turbulence level,  transport of particles and momentum, and the anisotropy in these quantities. In all the cases considered, turbulence amplitude is always quenched due to strong shear ($\xi =\nu k_y^2 / \A \ll 1$, where $\A$ is the shearing rate, $\nu$ is the molecular viscosity and $k_y$ is a characteristic wave-number of small-scale turbulence), with stronger reduction in the direction of the shear than those in the perpendicular directions. Specifically, in the large rotation limit ($\Omega \gg \A$), they scale as $\A ^{-1}$ and $\A^{-1} \vert \ln \xi \vert$, respectively, while in the weak rotation limit ($\Omega \ll \A$), they scale as  $\A^{-1}$ and $\A^{-2/3}$, respectively. Thus, flow shear always leads to weak turbulence with an effectively stronger turbulence in the plane perpendicular to shear than in the shear direction, regardless of rotation rate. The anisotropy in turbulence amplitude is however weaker by a factor of $\xi^{1/3} \vert \ln \xi \vert$ ($\propto \A^{-1/3} \vert \ln \xi \vert$) in the rapid rotation limit ($\Omega \gg \A$) than that in weak rotation limit ($\Omega \ll \A$) since rotation favors almost-isotropic turbulence. Compared to turbulence amplitude, particle transport is found to crucially depend on whether rotation is stronger or weaker than flow shear. When rotation is stronger than flow shear ($\Omega \gg \A$), the transport is inhibited by inertial waves, being quenched inversely proportional to the rotation rate (i.e. $\propto \Omega^{-1}$) while in the opposite case, it is reduced by shearing as $\A^{-1}$. Furthermore, the anisotropy is found to be very weak in the strong rotation limit (by a factor of 2) while significant in the strong shear limit. The turbulent viscosity is found to be negative with inverse cascade of energy as long as rotation is sufficiently strong compared to flow shear ($\Omega \gg A$) while positive in the opposite limit of weak rotation ($\Omega \ll \A$). Even if the eddy viscosity is negative for strong rotation ($\Omega \gg \A$), flow shear, which transfers energy to small scales, has an interesting effect by slowing down the rate of inverse cascade with the value of negative eddy viscosity decreasing as $|\nu_T| \propto \A^{-2}$ for strong shear. Furthermore, the interaction between the shear and the rotation is shown to give rise to a novel non-diffusive flux of angular momentum ($\Lambda$-effect), even in the absence of external sources of anisotropy. This effect provides a mechanism for the existence of shearing structures in astrophysical and geophysical systems. 
\end{abstract}

\pacs{47.27.Jv,47.27.T-,97.10.Kc}

\maketitle

\section{Introduction}
Rotating turbulent flows can be found in many areas such as engineering (turbo-machinery, combustion engine), geophysics (oceans, Earth's atmosphere) or astrophysics (gaseous planets, galactic and accretion disks). Large-scale fluid motions tend to appear as a robust feature in these systems, often in the form of shear flows (such as circulations on the surface of planets, differential rotation in stars and galaxies or flows in a rotating machinery). There have been accumulating evidence that large-scale shear flows as well as rotation play a crucial role in determining turbulence properties and transport, such as energy transfer or mixing (see below for more details). The understanding of the physical mechanism for the generation of large-scale shear flows and the complex interaction among rotation, shear flows and turbulence thus lies at the heart of the predictive theory of turbulent transport in many systems.

\subsection{Summary of previous works}
\label{Previous}
While both rotation and shear flow apparently have a similar effect on quenching turbulent transport, the efficiency of their effects as well as the basic physical mechanisms are totally different. It is thus useful to contrast these in detail.

\subsubsection{Sheared turbulence}
\begin{figure}
\begin{center}
\includegraphics[scale=0.7,clip]{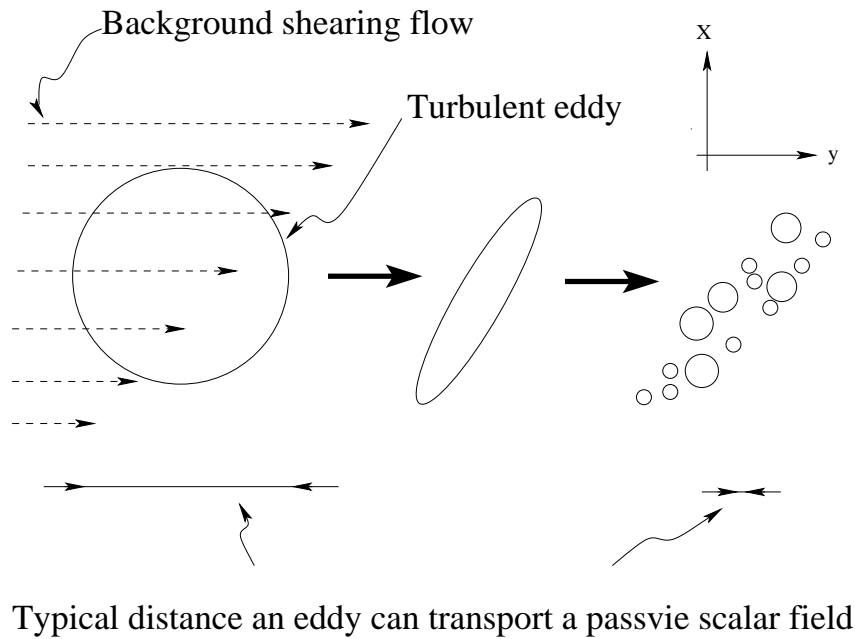}
\caption{\label{ShearEff} Sketch of the effect of shear on a turbulent eddy.}
\end{center}
\end{figure}

The main effect of shear flow is to advect turbulent eddies differentially, elongating and distorting their shapes, thereby rapidly generating small scales which are ultimately disrupted by molecular dissipation on small scales (see Fig. \ref{ShearEff}). That is, flow shear facilitates the cascade of various quantities such as energy or mean square scalar density to small scales (i.e. direct cascade) in the system, enhancing their dissipation rate. As a result, turbulence level as well as turbulent transport of these quantities can be significantly reduced compared to the case without shear. Another important consequence of shearing is to induce anisotropic transport and turbulent level since flow shear directly influences the component parallel to itself (i.e. $x$ component in Fig. 1) via elongation while only indirectly the other two components (i.e. $y$ and $z$ components in Fig. 1) through enhanced dissipation. This shearing effect of shear flow can be captured by time-dependent Fourier transform where the wave number in the shearing direction (e.g. $k_x$ in Fig. 1) increases linearly in time   \cite{Goldreich64,Townsend76,Kim05}.

It is important to emphasize that the aforementioned shearing effect (due to differential advection) is via nonlocal interaction between large and small-scale modes, and can dominate over nonlinear local interaction between small scales for sufficiently strong flow shear \cite[e.g.][]{Waleffe93}. Therefore, the evolution of small-scale quantities can be treated as linear by neglecting local interactions compared to nonlocal interactions. 
This formulation, also called the rapid distortion theory (RDT) by various previous authors \cite{Batchelor54,Townsend76}, was used to study the linear response of turbulence to a mean flow with spatially uniform gradients. The linear treatment of fluctuations by incorporating strong flow shear was also used in the astrophysical context by \cite{Goldreich64} by using shearing coordinates. The generation of large-scale shear flows (the so-called zonal flows) through a similar nonlocal interaction has been intensely studied in the magnetically confined plasmas, where turbulence quenching by shear flow is believed to be one of the most promising mechanisms for improving plasma confinement \cite{Burrell97,Kim04a}.  

In decaying sheared turbulence, \cite{Lee90} have shown a surprisingly good agreement between the RDT predictions and numerical simulations. Forced sheared turbulence was proposed for the first time by \cite{Nazarenko00a} in the context of two-dimensional near-wall turbulence to explain the logarithmic dependence of the large scale velocity on the distance to the wall. In that case, the external forcing is provided by a continuous supply of vorticity from intermittent coherent burst of vorticity coming from the viscous layer. This work was later generalized to three dimensions \cite{Nazarenko00b,Dubrulle01} with the same conclusions. Subsequently, theoretical predictions (using a quasi-linear theory) for the transport of passive scalar fields in 2D hydrodynamic turbulence by \cite{Kim03b} and \cite{Kim04b} have been beautifully confirmed by recent numerical simulations \cite{Leconte06}. In particular, they have shown that turbulent transport of particles can be severely quenched inversely proportional to flow shear $\A$ while turbulence level is reduced as $\A^{-5/3}$. Ref. \cite{Kim05} has shown that in 3D forced HD turbulence, strong flow shear can quench turbulence level and transport of particles with strong anisotropy (much weaker along the flow shear which is directly affected by shearing) and has emphasized the difference in turbulence level and transport, which is often used interchangeably in literature. A similar weak anisotropic transport was shown for momentum transport by \cite{2Shears} in forced 3D HD turbulence. Further investigations have been performed on turbulent transport in forced turbulence by incorporating the interaction of sheared turbulence with different types of waves that can be excited due to magnetic fields \cite{Kim01,Kim06,BetaPlane}, stratification \cite{Stratification} or both magnetic fields and stratification \cite{Dynamics}. 

\subsubsection{Rotating turbulence} 
Rotation has both similar and different effects on turbulent transport. First, rotation can reduce transport in the limit of rapid rotation (similarly to flow shear), but through a physical mechanism that is different from that of shear, namely by phase mixing of inertial waves \cite{Cally91}. It also induces only slight anisotropy in the transport (by a factor of two), much less significant than the strong anisotropy due to shear. Further, since phase mixing affects turbulent transport without necessarily quenching turbulence level, turbulence level may not be affected by rotation. This reduction in transport without much effect on turbulence level is a common feature of turbulence strongly affected by waves, and is also found in MHD turbulence where magnetic fields support Alfven waves \cite{Kim01,Cattaneo02,Kim06} and stratified turbulence \cite{Stratification} where stable stratification excites internal gravity waves. A more striking difference between flow shear and rotation is that rotation facilitates the cascade energy to large scale, generating large-scale flows. For instance, in the extreme limit of very rapid rotation, the fluid motion becomes independent of the coordinate along the rotation axis (the so-called Taylor-Proudman theorem \cite{Proudman16,Taylor21}). The generation of large-scale flow has been shown by various numerical simulations including \cite{Cambon97} and \cite{Smith99}. In particular, \cite{Smith99} have shown that the inverse cascade of energy is more pronounced in forced turbulence due to statistical triadic transfer through nonlocal interaction. 

It is important to note that this nonlocal interaction leading to inverse cascade can be successfully captured by inhomogeneous RDT theory which permits the feedback of the nonlinear local interaction between small scales onto the large scales via Reynolds stress (constituting the other part of quasi-linear analysis) while neglecting nonlinear local interaction between small scales for fluctuations compared to nonlocal interactions. As must be obvious by comparing the Coriolis force with nonlinear advection terms, the RDT works well for sufficiently strong rotation (small Rossby number) even in the absence of shear flow. For instance, the agreement of the RDT prediction with numerical results has been shown by various previous authors including \cite{Cambon97}, but mostly in decaying turbulence. However, in this case, the RDT cannot accurately capture the turbulence structure in the plane perpendicular to rotation axis where nonlinear local interactions between inertial waves seem important (see, e.g. \cite{Smith99}). The validity and weakness of the RDT together with comparison with various numerical simulation (without an external forcing) with/without shear flows and stratification can be found in excellent review by \cite{Salhi06} and Cambon and \cite{Salhi07b}, to which readers are referred for more details. 

In  comparison, far much less is understood in the case of forced turbulence. In particular, the main interest in forced turbulence is a long-term time behavior where the dissipation, enhanced by shear distortion, is balanced by energy input, thereby playing a crucial role in leading to a steady equilibrium state. The computational study of this long time behavior is however not only expensive but also difficult because of the limit on numerical accuracy, as noted by \cite{Salhi97}. Therefore, analytical theory by capturing shearing effect (such as quasi-linear theory with time-dependent wavenumber) would be extremely useful in obtaining physical insights into the problem as well as guiding future computational investigations. We note that the previous works by Kichatinov and Rudiger and collaborators \cite{Rudiger80,Kichatinov86b,Kichatinov87,Rudiger89,Kichatinov94} using quasi-linear theory are valid only in the limit of weak shear. We further note that physically, the local nonlinear interactions in Navier-Stokes equation can be captured by an external forcing \cite{Moffatt67,Cambon99}.


\subsection{Main objectives and methodology}
Our main motivation is to understand the origin of large-scale shear flow and its effect on turbulent transport in rotating systems. To this end, we consider a forced turbulence and investigate the combined effect of rotation and shear flow on the turbulence properties including transport of momentum and particles. Specifically, we are interested in how rotation and flow shear influence the generation of shear flow (e.g. the direction of energy cascade), turbulence level,  transport of particles and momentum, and the anisotropy in these quantities.
Given the differences/similarities in the effects of flow shear and rotation (as discussed in Sec. \ref{Previous}), of particular interest is to identify the relative strength of flow shear to rotation rate for the cross-over between inverse and direct cascades and isotropic and almost-isotropic turbulence/transport. Recalling that flow shear of strength $\A$ acts over the time-scale $\A^{-1}$ while rotation induces inertial waves of frequency $\sim \Omega$, one could naively think that flow shear would dominate the effect of rotation for sufficiently strong shear with $\A \gg \Omega$ while the effect of flow shear may be neglected in the opposite limit $\A \ll \Omega$. This will however be shown to be true only in the case of the transport of passive scalar fields and for the sign of eddy viscosity. That is, even in the case of weak shear compared to rotation $\A \ll \Omega$, the shear has yet a crucial effect on determining the overall amplitude of turbulence level and momentum transport since its shearing process (generating small scales) works coherently over more than one oscillation of the waves. To complement this, we are also interested in how shear-dominated turbulence is influenced by rotation. As will be shown later, when the system is linearly stable, weak rotation tends to make turbulence/transport more `isotropic'. 

Concerning momentum transport, another important question is the possibility of non-diffusive transport. In rotating turbulence, the inverse cascade can occur not only due to a (diffusive) negative viscosity, but also due to non-diffusive momentum transport. The latter is known as the anisotropic kinetic $\alpha$-effect (AKA) \cite{Frisch87} or as the $\Lambda$-effect in the astrophysical community. The appearance of non-diffusive term in the transport of angular momentum prevents a solid body rotation from being a solution of the Reynolds equation \cite{Lebedinsky41,Kippenhahn63}, and thus act as a source for the generation of large-scale shear flows. For instance, this effect has been advocated as a robust mechanism to explain the differential rotation in the solar convective zone. Starting from Navier-Stokes equation, it is possible to show that these fluxes arise when there is a cause of anisotropy in the system, either due to an anisotropic background turbulence (see \cite{Rudiger89} and references therein) or else due to inhomogeneities such as an underlying stratification. We will show that non trivial $\Lambda$-effect can result from an anisotropy induced by shear flow on the turbulence even when the driving force is isotropic, in contrast to the case without shear flow where this effect exists only for anisotropic forcing \cite{Kichatinov87}.

We note that although much less attention has been paid to the effect of rotation and shear on mixing and transport of scalars (such as pollutants, heat or reacting species) compared to momentum transport, this is an important problem in understanding the distribution and mixing of a variety of physical quantities in different systems. For instance, observations show that the concentration of light elements at the surface of the Sun is smaller than what is expected by comparison with Earth's or meteorites abundance. As these light elements can  only be destroyed below a strong shear layer (the so-called solar tachocline), their transport is subject to the effects of strong shear and rotation. The study of transport of passive scalar has been mostly limited to the purely rotating case \cite{Kaneda00,Cambon04} or non-rotating sheared turbulence \cite{Tavoularis81,Rogers89}. For purely rotating turbulence, linear theory has shown a strong suppression of particle diffusion by rotation, confirmed by numerical simulations \cite{Cambon04}. In comparison, the study of particle diffusion in sheared rotating turbulence was done only by \cite{Brethouwer05}, who found that numerical simulation results agree fairly well with his linear theory. 

The purpose of this paper is to provide theoretical prediction on these issues by considering a 3D incompressible fluid, forced by a small-scale external forcing. As we are interested in the effect of flow shear, we capture this effect non-perturbatively by using time-dependent wavenumber [see Eq. (\ref{TFdependent})]. By assuming either sufficiently strong shear or rotation rate, we employ a quasi-linear analysis to compute turbulence level, eddy viscosity, and particle transport for temporally short-correlated, homogeneous forcing. As the computation of these quantities involve too complex integrals to be analytically tractable, they are analytically computed by assuming an ordering in time scales. In our problem, there are three important (inverse) time-scales: the shearing rate $\A$, the rotation rate $\Omega$ and the diffusion rate $\mathcal{D}=\nu k_y^2$ where $\nu$ is the (molecular) viscosity of the fluid and $k_y^{-1}$ is a characteristic small scale of the system. We first distinguish the two cases of strong rotation ($\Omega \gg \A$) and weak rotation ($\Omega \ll \A$). The first regime of strong rotation will be studied in the strong shear ($\A \gg \mathcal{D}$) and weak shear ($\A \ll \mathcal{D}$) regime. On the other hand, the second regime of weak rotation will be considered only in the strong shear ($\A \gg \mathcal{D}$) case, as the effects of both shear and rotation disappear in the opposite limit ($\A \ll \mathcal{D}$). We believe that our results would provide not only useful physical insights in understanding the complex dynamics of rotating sheared turbulence, but also serve as a guide for further theoretical/computational works, especially considering the difficulty of numerical study of this system. 

The remainder of the paper is organized as follows: in \S \ref{Model}, we formulate our problem. Theoretical results of turbulent intensity and turbulent transport are provided in \S \ref{Equator}. Some of the detailed analysis are provided only in \S \ref{Equator}. We then discuss our findings in the strong shear limit in \S \ref{Implications} and provide concluding remarks in \S \ref{Conclusion}. The effect of rotation on linear stability of shear flows and some of the detailed algebra are provided in Appendices. Since analytical analysis performed in the paper are quite involved, some of the readers who are mainly interested in the results might wish to go to \S \ref{Implications} and \S \ref{Conclusion} after reading \S \ref{Model}.

\section{Model}
\label{Model}

We consider an incompressible fluid  in a rotating frame with average rotation rate $\tilde{\Omega}$, which are governed by
\EQA
\label{NSrotation}
\partial_t \uu + \uu \cdot \nabla \uu &=& - \nabla P + \nu \nabla^2 \uu + {\bf F} - 2 \tilde{\OOmega} \times \uu \; , \\ \nonumber
\nabla \cdot \uu &=& 0 \; .
\ENA

\begin{figure}
\begin{center}
\includegraphics[scale=1,clip]{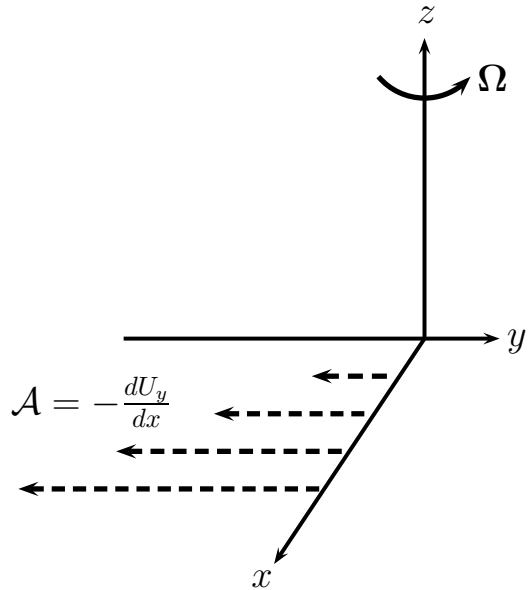}
\caption{\label{DessinEq} Sketch of the configuration in the perpendicular case}
\end{center}
\end{figure}
Following \cite{Kim05}, we study the effect of a large-scale shear $\UU_0 = U_0(x) \hat{j}$ on the transport properties of turbulence by writing the velocity as a sum of a shear (chosen in the $x$-direction) and fluctuations: $\uu = \UU_0 + \vv = U_0(x) \hat{j} + \vv = - x \mathcal{A} \hat{j} + \vv$. Without loss of generality, we assume $\A > 0$. In the following, we consider the configuration of Figure (\ref{DessinEq}) where the shear and rotation (in the $z$ direction) are perpendicular and  simplify notation by using ${\bf \Omega} = 2 \tilde{\bf \Omega}$. Then, the Coriolis force is simply
$\Omega [-u_y \, {\bf i} + u_x \, {\bf j}]$, where ${\bf i}$, ${\bf j}$  and ${\bf k}$  are the unit vectors associated with the Cartesian coordinates. Note that our $x-y$ coordinates are not conventional in that our $x$ and $y$ directions correspond to $y$ and $x$ in previous works (see \cite{Salhi97} for instance). Therefore, the shearing, the stream-wise and the span-wise direction correspond to the $x$, $y$ and $z$ direction, respectively.

To calculate turbulence amplitude (or kinetic energy) and turbulent transport, we need to solve the equation for the fluctuating velocity field. To this end, we employ the quasi-linear theory \cite{Moffatt78} where the nonlinear local interactions between small scales are neglected compared to nonlocal interactions between large and small scales and obtain:  
\EQA
\label{quasi-linear}
\partial_t \vv + \UU_0 \cdot \nabla \vv + \vv \cdot \nabla \UU_0  &=& - \nabla p + \nu \nabla^2 \vv + {\bf f} - {\OOmega} \times \vv \; , \\ \nonumber
\nabla \cdot \vv &=& 0 \; ,
\ENA
where $p$ and ${\bf f}$ are respectively the small-scale components of the pressure and forcing. As noted in the introduction, this approximation, also known as the RDT \cite{Townsend76}, is justified in the case of strong shear as the latter induces a weak turbulence, leading to weak interaction between small scales which is negligible compared to the (non-local) interaction between the shear and small scales. This has in fact been confirmed by direct numerical simulations, proving the validity of the predictions of quasi-linear theory with a constant-rate shear both in the non-rotating \cite{Lee90} and rotating unforced \cite{Salhi97} turbulence and also for forced turbulence \cite{Leconte06}. Further, note that the quasi-linear analysis is also valid in the limit of rapid rotation \cite{Cambon99}.

To solve Eq. (\ref{quasi-linear}), we introduce a Fourier transform with a wave number in the $x$ direction evolving in time in order to incorporate non-perturbatively the effect of the advection by the mean shear flow \cite{Goldreich64,Townsend76,Kim05}:
\begin{equation}
\label{TFdependent}
{\bf v}({\bf x},t) = \frac{1}{(2\pi)^2} \int d^3 k \; e^{i[{ k_x(t)} x + k_y y + k_z z]}
{\bf \tilde{v}}({\bf k},t) \; ,
\end{equation}
where $k_x(t) = k_x(0) + k_y \A t$. From Eqs. (\ref{quasi-linear}) and (\ref{TFdependent}), we obtain the following set of equations for the fluctuating velocity:
\EQA
\label{System1}
\A \partial_\tau \hat{v_x} &=& -i k_y \tau \hat{p} + \hat{f}_x + \Omega \hat{v}_y  \; ,\\ \nonumber
\A \partial_\tau \hat{v}_y -\A \hat{v_x} &=& -i k_y \hat{p} + \hat{f}_y - \Omega  \hat{v_x}  \; , \\ \nonumber
\A \partial_\tau \hat{v}_z &=& -i k_z \hat{p} + \hat{f}_z  \; , \\ \nonumber
0 &=& \tau \hat{v_x} + \hat{v}_y + \beta \hat{v}_z \; .
\ENA
Here, the new variables $\hat{\vv} = \tilde{\vv} \exp[\nu (k_H^2 t + k_x^3/3k_y\A)]$ and similarly for $\hat{\bf f}$ and $\hat{p}$ have been used to absorb the diffusive term, and the time variable has been changed to $\tau = k_x(t)/k_y$. In the remainder of the paper, we solve Eq. (\ref{System1}) for the fluctuating velocity (with a vanishing velocity as initial condition). We then use these results and the correlation of the forcing (defined in \S \ref{Forcing}) to compute the turbulence intensity and transport (defined in \S \ref{SectParticle}). 

\subsection{Transport of angular momentum}
\label{SectionEddy}
As the large-scale velocity is in the $y$ direction, we are mostly interested in the transport in that direction. The large-scale equation for the $y$ component of velocity $\UU_0$ is given by Eq. (\ref{NSrotation}) with a supplementary term ${\bf \nabla} \cdot {\bf R}$ where ${\bf R}$ is the Reynolds stress given by:
\EQ
{\bf R} = \langle {\bf v} v_y \rangle \; .
\EN

To understand the effect of ${\bf R}$ on the transport of angular momentum, one can formally Taylor expand it with respect to the gradient of the large-scale flow:
\EQ
\label{BullShit200}
R_i = \Lambda_i U_0 - \nu_T \partial_x U_0 \delta_{i1} + \dots = \Lambda_i U_0 + \nu_T \A \delta_{i1} + \dots\; .
\EN
Here, $\Lambda_i$ and $\nu_T$ are the two turbulent transport coefficients from non-diffusive and diffusive momentum flux, respectively. Note that the first term in the expansion is due to the small-scale driving and the Coriolis force in Eq. (\ref{NSrotation}) which break the Galilean invariance \cite{Dubrulle91}.
First, $\nu_T$ is the turbulent (eddy) viscosity, which simply changes the viscosity from the molecular value $\nu$ to the effective value $\nu + \nu_T$. Note that the sign of eddy viscosity represents the direction of energy cascade, with positive (negative) value for direct (inverse) cascade. Second, the first term involving $\Lambda_i$ in equation (\ref{BullShit200}) is proportional to the rotation rate rather than the velocity gradient. This means that it does not vanish for a constant velocity field and thus permits the creation of gradient in the large-scale velocity field. This term bears some similarity with the $\alpha$ effect in dynamo theory \cite{Parker55,Steenbeck66b} and has been known as the $\Lambda$-effect \cite{Lebedinsky41,Rudiger80} or anisotropic kinetic alpha (AKA)-effect \cite{Frisch87}. Similarly to the $\alpha$ effect, this effect exists only if the small-scale flow lacks parity invariance (going from right-handed to left handed coordinates). However, in contrast to the $\alpha$ effect, the $\Lambda$ effect requires anisotropy for its existence \cite{Rudiger80,Frisch87}. 

\subsection{Particle (or heat) transport}
\label{SectParticle}
To study the influence of rotation and shear on the particle and heat transport, we have to supplement equation (\ref{NSrotation}) with an advection-diffusion equation for these quantities. We here focus on the transport of particles since a similar result also holds for the heat transport. The density of particles $N({\bf x},t)$ is governed by the following equation:
\EQ
\partial_t N + \UU \cdot \nabla N = D \nabla^2 N \; ,
\EN
where $D$ is the molecular diffusivity of particle. Note that, in the case of heat equation, $D$ should be replaced by the molecular heat conductivity $\chi$. Writing the density as the sum of a large-scale component $N_0$ and small-scale fluctuations $n$ ($N = N_0 + n$), we can express the evolution of the transport of chemicals on large scales by:
\EQ
\partial_t N_0 + \UU_0 \cdot \nabla N_0 = (D \delta_{ij} + D_T^{ij}) \partial_i \partial_j N_0 \; ,
\EN
where the turbulent diffusivity is defined as $\langle v_i n \rangle = - D_T^{ij} \partial_j N_0$. $D_T^{ij}$ will analytically be computed to see the effect of rotation and flow shear on turbulent transport of chemicals which can be highly anisotropic. Note that the transport of a passive scalar quantity (contrary to the angular momentum which is a vector quantity) has to be diffusive due to the fact that it is solely advected by the flow \cite{Frisch89}.

For simplicity, we assume a unit Prandtl number $D = \nu$ and apply the transformation introduced in equation (\ref{TFdependent}) to the density fluctuation $n$ to obtain the following equation: 
\EQ
\label{EquTransport}
\partial_\tau \hat{n} = \frac{(- \partial_j N_0)}{\A} \hat{v}_j \; .
\EN
Equation (\ref{EquTransport}) simply shows that the fluctuating density of particles can be obtained by integrating the fluctuating velocity in time.

\subsection{External forcing}
\label{Forcing}
As mentioned in introduction, we consider a turbulence driven by an external forcing ${\bf f}$. To calculate the turbulence amplitude and transport defined in \S \ref{SectionEddy} and \S \ref{SectParticle} (which involve quadratic functions of velocity and/or density), we prescribe this forcing to be short correlated in time (modeled by a  $\delta$-function) and homogeneous in space with power spectrum $\psi_{ij}$ in the Fourier space. Specifically, we assume:
\begin{equation}
\label{ForcingCorrel}
\langle \tilde{f}_i({\bf k_1},t_1) \tilde{f}_j({\bf k_2},t_2) \rangle = \tau_f \, (2\pi)^3 \delta({\bf k_1}+{\bf k_2}) \, \delta(t_1-t_2) \, \psi_{ij}({\bf k_2}) \; ,
\end{equation}
for $i$ and $j$ = $1$, $2$ or $3$. The angular brackets stand for an average over realizations of the forcing, and $\tau_f$ is the (short) correlation time of the forcing. Note that the $\delta$ correlation is valid as long as the correction time $\tau_f$ is the shortest time-scale in the system [i.e. $\tau_f \ll \Omega^{-1}, \A^{-1}, 1/(\nu k^2)$].

For most results that will be derived later, we assume an incompressible and isotropic forcing where the spectrum of the forcing is given by:
\EQ
\label{ForcingIsotropic}
\psi_{ij}({\bf k}) = F(k) (\delta_{ij} - k_i k_j/k^2) \; .
\EN 
It is easy to check that in the absence of rotation and shear, this forcing leads to an isotropic turbulence with intensity:
\EQ
\label{IsotropicOriginel}
\langle v_{0}^2 \rangle = \frac{2 \tau_f}{(2 \pi)^2} \int_0^\infty \frac{F(k)}{\nu} \; dk \; ,
\EN
where the subscript $0$ stands for a turbulence without shear and rotation.

In addition to an isotropic forcing, we will also consider an anisotropic forcing in \S \ref{EqMomentWKB} to examine the combined effect of rotation and anisotropy, which can lead to non-diffusive fluxes of angular momentum. Specifically, we consider an extremely anisotropic forcing with motion restricted to a plane perpendicular to a given direction ${\bf g}$. The motion in this perpendicular plane is however assumed to be isotropic. Such a forcing can be modeled by the following power spectrum \cite{Rudiger89}:
\EQ
\label{ForcingAnisotropic}
\psi_{ij}({\bf k}) = G(k) \left[\delta_{ij} - \frac{k_i k_j}{k^2} - \frac{({\bf g} \cdot {\bf k})^2}{k^2} \delta_{ij} -  g_i g_j + \frac{{\bf g} \cdot {\bf k}}{k^2} (g_i k_j + g_j k_i) \right] \; .
\EN
In that case, the turbulence without rotation or shear would have the following properties:
\EQA
\label{anIsotropicOriginel}
\langle ({\bf v}_0 \cdot {\bf g})^2 \rangle &=&  0  \;  , \\ \nonumber
\langle ({\bf v}_0 \times {\bf g})^2 \rangle &=& \frac{2 \tau_f}{3(2 \pi)^2} \int_0^\infty \frac{G(k)}{\nu} \; dk \; .
\ENA

\section{Analytical results}
\label{Equator}
The system (\ref{System1}) can be simplified to:
\EQA
\label{System3}
\partial^2_\tau \bigl[(\gamma + \tau^2) \hat{v_x} \bigr] &+&  \beta^2 \OB(\OB-1) \hat{v_x} = \partial_\tau \Bigl[\frac{h_1(\tau)}{\A}\Bigr] - \bar{\Omega} \beta \frac{h_2(\tau)}{\A} \; , \\ \nonumber
\partial_\tau \hat{v}_z &=& - \frac{\beta}{\gamma} \partial_\tau \bigl[ \tau \hat{v_x} \bigr] + \beta \frac{\OB-1}{\gamma} \hat{v_x} + \frac{h_2(\tau)}{\gamma \A} \; , \\ \nonumber
\hat{v}_y &=& - (\tau \hat{v_x}  + \beta \hat{v}_z) \; .
\ENA
Here:
\EQA
\label{Bullshit500}
\OB &=& \Omega / \A \; , \quad  \beta = k_z/k_y \; , \quad  \gamma = 1+\beta^2 = k_H^2 / k_y^2 \quad (k_H^2 = k_y^2 + k_z^2) \; , \\ \nonumber
h_1(\tau) &=& \gamma \hat{f}_x - \tau \hat{f}_y - \beta \tau \hat{f}_z \; , \quad
h_2(\tau) = \hat{f}_z - \beta \hat{f}_y \; .
\ENA
To solve the first of equation (\ref{System3}) which is a non-homogeneous second order differential equation, we need two boundary conditions. We impose a vanishing initial velocity ${\bf v}(\tau_0) =  0$  which implies $\hat{v}_x(\tau_0) = 0$ and $\partial_\tau \hat{v_x} \vert_{\tau=\tau_0} = h_1(\tau_0) / (\gamma + \tau_0^2) \A$. The second boundary condition can be shown to be obtained in the intermediate steps of deriving Eq. (\ref{System3}).

The exact solution to (\ref{System3}) is obtained in the appendix \ref{LinStab}, where we address the stability of the homogeneous solution of system (with ${\bf f = 0}$). Computations of correlation functions, by using this exact solution, however turns out to be too complex to be analytically tractable. To gain a physical insight into the role of inertial waves and flow shear in turbulent transport, we  consider the two limits -- (i) the strong rotation  where the effect of waves dominates shearing ($\Omega \gg \A$) and (ii) the weak rotation where shearing dominates the effects of waves ($\Omega \ll \A$). Approximate solutions can be derived in these two regimes which can then be used for deriving analytic form of correlation functions for turbulence intensity and transport. 

\subsection{Rapid rotation limit: $\Omega \gg \A$}
\label{EqLargeRotation}
When the rotation rate is much larger than the shearing rate ($\OBN = \vert \Omega \vert / \A \gg 1$), the oscillation of inertial waves is roughly coherent without being damped over shearing time of $\A^{-1}$. Therefore, these waves can play a dominant role in determining the direction of energy cascade (sign of eddy viscosity) and transport of particles via phase mixing (i.e. by affecting the phase relation). However, as shown below, flow shear can still have a non-trivial effect on turbulence by enhanced dissipation so long as it is stronger than molecular dissipation. To characterize the latter, we introduce a parameter $\xi = \nu k_y^2 /\A$, the ratio of typical molecular dissipation rate to shearing rate. Here, $k_y$ is the characteristic wavenumber of the forcing in the stream-wise direction. We can, for instance, envision the forcing to have a spectrum peaked around this characteristic wave-number $k_y$. In the following, we examine the changes in turbulence characteristics in weak ($\xi \gg 1$) and strong ($\xi \ll 1$) shear limits to elucidate the effects of flow shear in inertial wave-dominated turbulence. 

In the rapid rotation limit ($\vert \Omega \vert \gg \A$), the solution of Eq. (\ref{System3}) can be found by using WKB approximation \cite{Bender} as:
\EQA
\nonumber
\hat{v}_x(\tau) &=& \frac{1}{\A (\gamma + \tau^2)^{3/4}} \int_{\tau_0}^\tau dt \left\{ \frac{\hat{h}_1(t)}{(\gamma+t^2)^{1/4}} \cos[v(t,\tau)] + \hat{h}_2(t) (\gamma+t^2)^{1/4} \theta \sin[v(t,\tau)] \right\} \; , \\ \nonumber
\hat{v}_y(\tau) &=& \frac{1}{\A \gamma (\gamma + \tau^2)^{3/4}} \int_{\tau_0}^\tau dt \Bigl\{ \frac{\hat{h}_1(t)}{(\gamma+t^2)^{1/4}} \left(-\tau \cos[v(t,\tau)] + \beta \theta \sqrt{\gamma+\tau^2} \sin[v(t,\tau)] \right)  \\  \label{WKBEquator}
&& \qquad + \hat{h}_2(t) (\gamma+t^2)^{1/4} \left(-\theta \tau \sin[v(t,\tau)] -\beta \sqrt{\gamma+\tau^2} \cos[v(t,\tau)] \right) \Bigr\} \; , \\ \nonumber
\hat{v}_z(\tau) &=& \frac{1}{\A \gamma (\gamma + \tau^2)^{3/4}} \int_{\tau_0}^\tau dt \Bigl\{ \frac{\hat{h}_1(t)}{(\gamma+t^2)^{1/4}} \left(- \beta \tau \cos[v(t,\tau)] - \theta \sqrt{\gamma+\tau^2} \sin[v(t,\tau)] \right) \\ \nonumber
&& \qquad + \hat{h}_2(t) (\gamma+t^2)^{1/4} \left(-\theta \beta \tau \sin[v(t,\tau)] + \sqrt{\gamma+\tau^2} \cos[v(t,\tau)] \right) \Bigr\} \; .
\ENA
Here,
\EQA
\OBN &=& \vert \OB \vert \quad , \qquad \omega_0 = \vert \beta \vert \OBN \quad , \qquad \theta = \text{sign}(\beta \OB) \; , \\ \nonumber
s(t) &=& \left(1-\frac{1}{2 \OB}\right) \text{arcsinh}\left(\frac{t}{\sqrt{\gamma}}\right) + O\left(\frac{1}{\OBN^2}\right) \; , \\ \nonumber
v(t,\tau) &=& \omega_0 \left[s(t) - s(\tau)\right] \; .
\ENA 
In the following subsections, we compute the various correlation functions by assuming a homogeneous and short-correlated forcing [see Eq. (\ref{ForcingCorrel})]. As the system (\ref{System3}) involves the forcing in terms of  $\hat{h}_1$ and $\hat{h}_2$ only [see Eq. (\ref{Bullshit500})], it is convenient to use the power spectrum $\phi_{ij}$ as:
\begin{equation}
\label{ForcingCorrelbis}
\langle \tilde{h}_i({\bf k_1},t_1) \tilde{h}_j({\bf k_2},t_2) \rangle = \tau_f \, (2\pi)^3 \delta({\bf k_1}+{\bf k_2}) \, \delta(t_1-t_2) \, \phi_{ij}({\bf k_2}) \; ,
\end{equation}
for $i$ and $j$ = $1$ or $2$. In the case of an isotropic and incompressible forcing [Eq. (\ref{ForcingIsotropic})], $\phi_{ij}$ in Eq. (\ref{ForcingCorrelbis}) can be written:
\EQA
\label{PhiIsotropic}
\phi_{11}({\bf k}) = \gamma (\gamma+a^2) F(k) \; ,
\phi_{12}({\bf k}) = 0 \; ,
\phi_{22}({\bf k}) = \gamma F(k) \; .
\ENA

\subsubsection{Turbulence intensity}
We begin by examining the effects of rotation and flow shear on turbulence level in wave-dominated turbulence due to strong rotation ($\vert \Omega \vert \gg \A$). The effect of shear will further be clarified by comparing results in weak shear limit ($\xi \gg 1$) with those in the strong shear limit ($\xi \ll 1$). First, turbulence intensity in the shear direction can be obtained by using Eqs (\ref{WKBEquator}) and (\ref{ForcingCorrelbis}) as:
\EQA
\label{Vx1}
\langle v_x^2 \rangle &=& \frac{\tau_f}{(2 \pi)^3 \A} \int d^3 k \int_a^{+\infty} d\tau \frac{e^{-2\xi \left[Q(\tau)-Q(a)\right]}}{(\gamma+\tau^2)^{3/2}}  \Bigl\{ \frac{\phi_{11}({\bf k})}{\sqrt{\gamma+a^2}} \cos^2[v(a,\tau)] \\ \nonumber
&& \qquad + \theta \phi_{12}({\bf k}) \sin[2 v(a,\tau)] + \phi_{22}({\bf k}) \sqrt{\gamma+a^2} \sin^2[v(a,\tau)] \Bigr\}\; .
\ENA
Here, $a=k_x / k_y$, $\beta = k_z/k_y$, $\gamma = 1+\beta^2$, $\xi = (\nu k_y^2)/\A$ and $Q(x) = x^3 / 3 + \gamma x$. In the case of an isotropic forcing [Eq. (\ref{PhiIsotropic})], Eq. (\ref{Vx1}) and the turbulence intensity in the two other directions can then be derived as:
\EQA
\label{TurbulentIntensity}
\langle v_x^2 \rangle &=& \frac{\tau_f}{(2 \pi)^3 \A} \int d^3 k \; \gamma \sqrt{\gamma + a^2} F(k) \; I^0({\bf k}) \; , \\ \nonumber
\langle v_y^2 \rangle &=& \frac{\tau_f}{(2 \pi)^3 \A} \int d^3 k \; \sqrt{\gamma + a^2} F(k) \; \left\{\beta^2 I^0{\bf k}) + I^2({\bf k}) \right\} \; ,\\ \nonumber
\langle v_z^2 \rangle &=& \frac{\tau_f}{(2 \pi)^3 \A} \int d^3 k \; \sqrt{\gamma + a^2} F(k) \; \left\{I^0({\bf k}) + I^2({\bf k}) \right\} \; .
\ENA
Here:
\EQ
\label{Ip}
I^p({\bf k}) = \int_a^{+\infty}  \frac{\tau^p \, e^{-2\xi \left[Q(\tau)-Q(a)\right]}}{(\gamma+\tau^2)^{3/2}}  d\tau \; .
\EN
In order to understand the effect of shearing on turbulence intensity in this wave-dominated turbulence, we first examine (\ref{TurbulentIntensity}) in the weak shear limit ($\xi \gg 1$) where the shear is negligible. In this case, the integral $I_p$ in Eq. (\ref{Ip}) takes the approximate value:
\EQ
\label{Iplarge}
I^p({\bf k}) \sim \frac{a^p}{2 \xi (\gamma+a^2)^{5/2}} = \frac{\A a^p}{2 \nu k^2 (\gamma+a^2)^{3/2}} \; . 
\EN

By using Eq. (\ref{Iplarge}) in Eq. (\ref{TurbulentIntensity}), we can then obtain the following result for the turbulent intensity:
\EQA
\label{TurbulenceWeakShear}
\langle v_x^2 \rangle &=& \frac{\tau_f}{(2 \pi)^3} \int d^3 k \; \frac{F(k)}{2 \nu k^2} \frac{\gamma}{\gamma+a^2} \; , \\ \nonumber
\langle v_y^2 \rangle &=& \frac{\tau_f}{(2 \pi)^3} \int d^3 k \; \frac{F(k)}{2 \nu k^2} \frac{\beta^2+a^2}{\gamma+a^2} \; ,\\ \nonumber
\langle v_z^2 \rangle &=& \frac{\tau_f}{(2 \pi)^3} \int d^3 k \; \frac{F(k)}{2 \nu k^2} \frac{1+a^2}{\gamma+a^2} \; .
\ENA
Performing the integration over the angular variable, we obtain:
\EQA
\nonumber
\langle v_x^2 \rangle &=& \frac{\tau_f}{(2 \pi)^3} \int d k \; \frac{F(k)}{2 \nu } \int_0^{2\pi} d \phi \int_0^\pi d \theta \sin\theta \left(\cos^2 \theta + \sin^2\theta \sin^2\phi \right) \\  
\label{Bullshit600}
&=& \frac{2 \tau_f}{3 (2 \pi)^2} \int d k \; \frac{F(k)}{\nu } 
=\frac{1}{3}\langle v_0^2 \rangle\; , \\
\nonumber 
 \langle v_y^2 \rangle &=&\langle v_z^2 \rangle 
=\frac{1}{3}\langle v_0^2 \rangle\; . 
\ENA
Here, $\langle v_0^2 \rangle$ is the turbulence amplitude in the absence of rotation and shear [see Eq. (\ref{IsotropicOriginel})]. These results thus show that, in the large rotation limit, the turbulence intensity is isotropic and equals to the one without rotation [see Eq. (\ref{IsotropicOriginel})] for sufficiently weak shear with $\xi \gg 1$. Furthermore, in this limit of a sufficiently weak shear where $(\Omega, \mathcal{D}) \gg \A$, turbulence intensity is independent of rotation since waves do not necessarily quench turbulence level. A similar result was also obtained in MHD turbulence and stratified turbulence where magnetic fields and gravity waves mainly affect transport without much effect on turbulence level \cite{Kim06,Dynamics,Stratification}. We shall show below that a strong anisotropy can be induced when shearing effect is not negligible ($\xi \ll 1$) even in the rapid rotation limit ($\Omega \gg \A$). 

In order to understand the effect of  flow shear, we now consider the strong shear limit ($\xi \ll 1$). In this limit, the integral (\ref{Ip}) is simplified as:
\EQA
\label{Ipweak}
I^0({\bf k}) &=& \frac{1}{\gamma} \left(1 - \frac{a}{\sqrt{\gamma+a^2}} \right) \; , \\ \nonumber
I^2({\bf k}) &=& \frac{-\ln\xi}{3} \; .
\ENA
By plugging Eq. (\ref{Ipweak}) in Eq. (\ref{TurbulentIntensity}), we obtain:
\EQA
\label{TurbulenceStrongShear}
\langle v_x^2 \rangle &=& \frac{\tau_f}{(2 \pi)^3 \A} \int d^3 k \; \sqrt{\gamma + a^2} F(k)  
\;\;\propto \; \xi \langle v_0^2 \rangle
\; , \\ \nonumber
\langle v_y^2 \rangle &=& \frac{\tau_f}{(2 \pi)^3 \A} \int d^3 k \; \sqrt{\gamma + a^2} F(k) \frac{-\ln\xi}{3} 
\;\;\propto \; \xi |\ln \xi|\langle v_0^2 \rangle
\; ,\\ \nonumber
\langle v_z^2 \rangle &=& \frac{\tau_f}{(2 \pi)^3 \A} \int d^3 k \; \sqrt{\gamma + a^2} F(k) \; \frac{-\ln\xi}{3} 
 \;\;\propto \; \xi |\ln \xi|\langle v_0^2 \rangle
\; ,
\ENA
to leading order in $\xi \ll 1$. Note that in the calculation of $\langle v_x^2 \rangle$, we neglected the component proportional to $a=k_x / k_y$ as it is odd in both $k_x$ and $k_y$ and thus vanishes after integration over the angular variables for an isotropic forcing. The last terms in Eq. (\ref{TurbulenceStrongShear}), expressed in terms of the turbulence amplitude in the absence of rotation and shear $\langle v_0^2 \rangle$ [see Eq. (\ref{IsotropicOriginel})], explicitly show the dependence of turbulence level on rotation and shear. That is, all the components of turbulence intensity is reduced for strong shear $\xi \ll 1$. Further, the $x$ component along shear is reduced as $\xi \propto \A^{-1}$ while the other two components as $\xi |\ln \xi|$, with an effectively weaker turbulence in the shear direction than in the perpendicular one, by a factor of $\ln\xi$. This shows that shear flow can induce anisotropic turbulence (unlike rotation) even when the forcing is isotropic. This result is similar to that obtained in the simulation of a Couette flow at high rotation rate \cite{Bech97} where the velocity fluctuations perpendicular to the wall exceed that in the stream-wise direction. Nevertheless, Eq. (\ref{TurbulenceStrongShear}) shows that a strong rapid rotation yet insures an isotropy in velocity fluctuations in $y-z$ directions ($\langle v_y^2 \rangle = \langle v_z^2 \rangle$). 


\subsubsection{Transport of angular momentum}
\label{EqMomentWKB}
As noted in the Introduction, rotation tends to cascade energy to large scales while shear flow to small scales. Would thus the inverse cascade be a robust feature for rapid rotation ($\Omega \gg \A$)? If yes, what would the effect of flow shear? Would there be a non-diffusive momentum transport? We answer these questions by first considering an isotropic forcing and then anisotropic forcing. The effect of shear will be elucidated by looking at the two limits of weak shear ($\xi \gg 1$) and strong shear ($\xi \ll 1$), as done in \S 3.1.1.

First, in the case of an isotropic forcing, we obtain the following Reynolds stress from equations (\ref{WKBEquator}) and (\ref{ForcingCorrelbis}):
\EQ
\label{TurbulentViscosity}
\langle v_x v_y \rangle = - \frac{\tau_f}{(2 \pi)^3 \A} \int d^3 k \;  \sqrt{\gamma + a^2} F(k) \; I^1({\bf k}) \; ,
\EN
where $I^1$ was defined in Eq. (\ref{Ip}). Eq. (\ref{TurbulentViscosity}) is computed in the weak and strong shear limits, below.

First, in the weak shear limit ($\xi \gg 1$), there is no contribution to leading order in $\OB^{-1}$ as the function $I^1$ is odd in $a$ and thus vanishes after integration over the wave vector. We thus include one higher order in $\OB^{-1}$ in the expansion and obtain the following result:
\EQ
\label{EV2}
\langle v_x v_y \rangle = - \frac{\tau_f}{(2 \pi)^3 \A} \int d^3 k \;  \frac{a  F(k)}{2 \omega_0} \; J({\bf k}) \; .
\EN
Here, we defined a function $J({\bf k})$, which has the following asymptotic behavior in the weak shear limit:
\EQA
\label{Jweak}
J({\bf k}) &=& \int_a^{+\infty}  \frac{\tau e^{-2\xi \left[Q(\tau)-Q(a)\right]}}{(\gamma+\tau^2)^{3/2}}  \sin\left[2 \omega_0 \left\{ s(a)-s(\tau) \right\} \right] d\tau \\ \nonumber
&\sim&  - \frac{ a \overline{\omega_0} \A}{2(\gamma+a^2)^{3/2} [\nu^2 k^4 + \overline{\omega_0}^2]} \; ,
\ENA
where $\overline{\omega_0} = \omega_0 \A / \sqrt{\gamma+a^2}$. Plugging Eq. (\ref{Jweak}) in Eq. (\ref{EV2}) and performing the integration over the azimuthal angle variable $\phi$, we obtain:
\EQ
\label{Bullshit200}
\langle v_x v_y \rangle = \frac{\tau_f \A}{32 (3 \pi)^2} \int d k \; k^2 F(k) \int_0^\pi d\theta \sin^5\theta \frac{1}{\nu^2 k^4 + \overline{\omega_0}^2} \; .
\EN
Finally, we change the integration variable from $\theta$ to $\overline{\omega_0} = \Omega \cos\theta$, obtaining the following formula:
\EQ
\langle v_x v_y \rangle = \frac{\tau_f \A}{16 (2 \pi)^2 \vert \Omega \vert} \int_0^{+\infty} d k \; k^2 F(k) \int_0^{\vert \Omega \vert} d\overline{\omega_0} \frac{\left(1-\overline{\omega_0}^2/\Omega^2\right)^2}{\nu^2 k^4 + \overline{\omega_0}^2} \; .
\EN
Therefore, in the large rotation and weak shear limit, the Reynolds stress becomes purely diffusive (with no $\Lambda$-effect) with the turbulent viscosity:
\EQ
\label{nutWKBequatorWeak}
\nu_T \sim  \frac{\pi \tau_f}{32 (2 \pi)^2 \vert \Omega \vert} \int_0^{+\infty} d k \frac{F(k)}{\nu}   \; .
\EN
This result shows that the turbulent viscosity is positive and proportional to $\Omega^{-1}$ for large $\Omega$. It is worth comparing Eq. (\ref{nutWKBequatorWeak}) with Eq. (22)  in  \cite{Kichatinov86b}. To this end, we use Eq. (\ref{IsotropicOriginel}), which gives the turbulence amplitude without rotation (the original turbulence of Kichatinov) in Eq. (\ref{nutWKBequatorWeak}) to obtain the turbulent viscosity  $\nu_T \sim \pi \langle v_{0}^2 \rangle / 64 \vert \Omega \vert$. Thus $\nu_T$ in Eq. (\ref{nutWKBequatorWeak}) is the same as  Eq. (22)  in  \cite{Kichatinov86b}  for $\vert \Omega \vert \gg 1$ and $\theta = \pi /2$, but has an opposite sign. This is due to the $\tau$-approximation used by Kichatinov which gave an unphysical result. Later, \cite{Kichatinov88} showed that the viscosity is also positive at any rotation rate  when derived consistently with quasi-linear approximation in the weak shear limit.

In comparison, in the strong shear limit ($\xi \ll 1$), the function $I^1$ in Eq. (\ref{Ip}) has the following asymptotic behavior:
\EQ
\label{I1weak}
I^1({\bf k}) = \frac{1}{\sqrt{\gamma+a^2}} \; .
\EN
Plugging Eq. (\ref{I1weak}) in Eq. (\ref{TurbulentViscosity}), we obtain the turbulent viscosity in the strong shear limit as:
\EQ
\label{nutWKBequator}
\nu_T = \frac{\langle v_x v_y \rangle}{\A} = - \frac{\tau_f}{(2 \pi)^3 \A^2} \int d^3 k \;   F(k)  \; .
\EN
Eq. (\ref{nutWKBequator}) shows that the turbulent viscosity is negative (as $F(k) > 0$) in the strong shear limit, in sharp contrast to the weak shear limit where $\nu_T > 0$ [see Eq. (\ref{nutWKBequatorWeak})]. Furthermore, the magnitude of $\nu_T$ is reduced by the shear ($\propto \A^{-2}$) and is independent of rotation, which should also be compared with the weak shear limit [see Eq. (\ref{nutWKBequatorWeak}) where $\nu_T \propto \Omega^{-1}$]. Therefore,  the turbulent viscosity changes from positive (for weak shear) to negative (for large shear) as the ratio of shear to dissipation increases. This result can be understood if we assume that, as in most rapidly rotating fluid, the inverse cascade is associated with the conservation of a potential vorticity \cite{Pedlovsky}. In the presence of strong shear (compared to dissipation), the potential vorticity is strictly conserved giving rise to an inverse cascade (negative viscosity). When the dissipation increases, the potential vorticity is less and less conserved and thus the inverse cascade is quenched. Our results show that there is a transition from inverse to direct cascade as the dissipation is increased. A similar behavior is also found in two-dimensional hydrodynamics (HD) where an inverse cascade can be shown to be present only for sufficient weak dissipation \cite{Kim01}.

It is important to note that the negative viscosity $\nu_T<0$ obtained here for strong rotation/strong shear ($\Omega \gg \A \gg \nu k_y^2$) signifies the amplification of shear flow as the effect of rotation favoring inverse cascade dominates shearing (generating small scales). However, the magnitude of $\nu_T$ is reduced by shear as $|\nu_T| \propto \A^{-2}$ since flow shear  inhibits the inverse cascade. This can be viewed as `self-regulation' -- that is, self-amplification of shear flow is slowed down as the latter becomes stronger.

The preceding results [Eqs. (\ref{nutWKBequatorWeak}) and (\ref{nutWKBequator})] indicate that in the large rotation limit where rotation dominates over shear, the momentum transport is purely diffusive for isotropic forcing, with opposite sign of turbulent viscosity for weak ($\xi \gg §1$) and strong shear ($\xi \ll §1$) for a fixed value of $\vert \Omega \vert / \A$ ($\gg 1$). In the case of anisotropic forcing, there is however a possibility of the appearance of non-diffusive momentum transport ($\Lambda$-effect). To examine this possibility, we now consider an extremely anisotropic forcing (introduced in \S \ref{Forcing}) where the forcing is restricted to horizontal plane ($y$-$z$), perpendicular to the direction of the shear. Using Eq. (\ref{ForcingAnisotropic}) with $g_{ij} = \delta_{i1}$, we obtain the following Reynolds stress: 
\EQ
\label{Viscosite}
\langle v_x v_y \rangle = - \frac{\tau_f}{(2 \pi)^3 \A} \int d^3 k \;  \frac{\gamma  G(k)}{2 \sqrt{\gamma+a^2}} \; \left[\left\{I^1({\bf k})-J'({\bf k})\right\} + \beta \theta K({\bf k}) \right]  \; .
\EN
Here, $I^1$ was defined previously in Eq. (\ref{Ip}) and:
\EQA
J'({\bf k}) = \int_a^{+\infty}  \frac{\tau e^{-2\xi \left[Q(\tau)-Q(a)\right]}}{(\gamma+\tau^2)^{3/2}}  \cos\left[2 \omega_0 \left\{ s(a)-s(\tau) \right\} \right] d\tau  \; , \\ \nonumber
K({\bf k}) = \int_a^{+\infty}  \frac{e^{-2\xi \left[Q(\tau)-Q(a)\right]}}{(\gamma+\tau^2)}  \sin\left[2 \omega_0 \left\{ s(a)-s(\tau) \right\} \right] d\tau  \; .
\ENA 
 
We again consider the weak and strong shear limits in the following. First, in the weak shear limit ($\xi \gg 1$), Eq. (\ref{Viscosite}) is simplified to:
\EQ
\label{Bullshit1}
\langle v_x v_y \rangle = \frac{\tau_f}{(2 \pi)^3 \A} \int d^3 k \;  \frac{\gamma  G(k) \beta \theta}{4 (\gamma+a^2)^{3/2}} \; \frac{\overline{\omega_0}}{\nu^2 k^4 + \overline{\omega_0}^2} \; .
\EN
Performing the angular integration in Eq. (\ref{Bullshit1}) and taking the large rotation limit, we obtain the following:
\EQ
\label{Bullshit2}
\langle v_x v_y \rangle = \frac{\tau_f}{3 (2 \pi)^3 \Omega \A} \int d^3 k \;  \frac{G(k)}{\nu}  \; .
\EN
Equation (\ref{Bullshit2}) is odd in the rotation and thus represents the $\Lambda$-effect. Again, the latter favors the creation of velocity gradient rather than smoothing it out and can thus provide a mechanism for the occurrence of differential rotation (e.g., in the sun). By using Eq. (\ref{anIsotropicOriginel}), one can see that the $\Lambda$-effect is proportional to the anisotropy in the turbulence without shear and rotation. This result shows that, in the large rotation limit, one needs anisotropic forcing to generate non-diffusive fluxes of angular momentum \cite[as in the case without shear as shown][]{Kichatinov86b}. This should be contrasted to the case of weak rotation (see \S \ref{WeakRotEq}) where the shear can alone give rise to an anisotropic turbulence, thereby leading to a $\Lambda$-effect even with an isotropic forcing.

Finally, in the opposite, strong shear limit ($\xi \ll 1$), Eq. (\ref{Viscosite}) becomes:
\EQ
\langle v_x v_y \rangle = - \frac{\tau_f}{(2 \pi)^3 \A} \int d^3 k \;  \frac{\gamma  G(k)}{2 (\gamma+a^2)}  \; ,
\EN
which is even in the rotation. Thus, the turbulent viscosity $\nu_T$ is obviously negative. Thus, in the large shear limit (but still negligible compared to the rotation), anisotropic forcing does not induce any non-diffusive fluxes but just increases the magnitude of the negative turbulent viscosity.

\subsubsection{Transport of particles}
In the large rotation limit ($\vert \Omega \vert / \A \gg 1$), inertial waves might play a crucial role in transport of particle as waves can alter the phase relation between particle density and velocity, as noted previously. How does this effect appear in forced turbulence? What is the effect of shear flow on particle transport dominated by waves? These questions are answered in this subsection.

In the rapid rotation limit ($\vert \Omega \vert / \A \gg 1$), turbulent particle diffusivities can be obtained after a long, straightforward analysis (see Appendix \ref{Appendix2} for details about the algebra) as: 
\EQA
\label{Isotropic2}
D_T^{xx} &\sim& \frac{\tau_f}{8 \pi \vert \Omega \vert} \int_0^{\infty} \frac{F(k)}{\nu}  \; dk \; , \\ \nonumber
D_T^{yy} = D_T^{zz} &\sim& \frac{\tau_f}{16 \pi \vert \Omega \vert} \int_0^{\infty} \frac{F(k)}{\nu} \; dk  \;\sim \; \frac{1}{2} D_T^{xx}\; .
\ENA
Note that in that case, the result is not sensitive to the value of the parameter $\xi$ and thus we do not distinguish between the weak and large shear limits. Eq. (\ref{Isotropic2}) shows that $D_T^{xx}$, $D_T^{yy}$ and $D_T^{zz}$ are all reduced as $\Omega^{-1}$ (with no effect of the shear) for large $\Omega$ and also that there is only a slight anisotropy in the transport of scalar: the transport in the direction of the rotation is twice larger than the one in the perpendicular direction \cite{Kichatinov94}. Interestingly, this anisotropy in the transport of particles is not present in turbulence intensity [see Eq. (\ref{Bullshit600})]. This is because waves can affect the phase between density fluctuation and velocity, not necessarily altering their amplitude. However, it is important to note that this anisotropy is only a factor of 2, much weaker than that in sheared turbulence without rotation \cite{Kim05}.

To summarize, in this subsection 3.1, we have examined how a shear flow can affect the turbulent property when turbulence is largely dominated by inertial waves in rapid rotation limit ($\vert \Omega \vert / \A \gg 1$). In particular, the results show:
\begin{enumerate}
\item that shear flow reduces turbulence level with a strong anisotropy [Eq. (\ref{TurbulenceStrongShear})], leading to an effectively weaker turbulence in the direction of the shear [which would otherwise be almost isotropic [Eq. (\ref{TurbulenceWeakShear})]; 
\item that in comparison, transport of particles is mainly governed by waves with almost isotropic property (within a factor of 2) and quenched as $\Omega^{-1}$ as rotation rate $\Omega$ increases; \item that energy cascade is inverse with negative viscosity for strong rotation/shear limit ($\Omega \gg \A \gg \nu k_y^2$) while its rate is slowed down by strong shear; 
\item that momentum transport is purely diffusive for isotropic forcing, with non diffusive transport appearing only for anisotropic forcing.
\end{enumerate}

\subsection{Weak rotation limit: $\Omega \ll \A$}
\label{WeakRotEq}
When $\Omega \ll \A$, flow shear can distort inertial waves over the period of their oscillation, dramatically weakening the effects of these waves on turbulence. Therefore, shear may take a dominant role in determining turbulence property (studied in \cite{Kim05}) while rotation modifies some of the properties of this shear-dominated turbulence. The investigation of this limit would thus permit us to clarify the effects of rotation as well as flow shear, thereby complementing the analysis done in Sec. \ref{EqLargeRotation} for strong rotation ($\Omega \gg \A$). Of particular interest is (1) to what extent the quenching and anisotropy of sheared turbulence \cite{Kim05} are affected by rotation, which favors isotropic turbulence; (2) how the direction of the energy cascade, which tends to be direct in 3D sheared turbulence, is affected by rotation (which prefers inverse cascade); (3) whether momentum transport can occur via non-diffusive fluxes.

To answer these questions, we expand various physical quantities in powers of $\OBN = \vert \Omega \vert / \A$ as:
\EQ
\label{ExpansionX}
X(\tau) = X_0(\tau) + \OBN X_1(\tau) + \dots \; ,
\EN
in the weak rotation limit ($\Omega \ll \A$) and calculate the turbulence intensity and transport up to first order in $\OBN$. For the sake of brevity, we here just provide the final results of the calculation. Note that in this limit, we are only interested in strong shear case ($\xi \ll 1$) since in the opposite limit where $\nu k_y^2 \gg \A \gg \Omega$, the effects of both shear and rotation simply disappear to leading order. 

\subsubsection{Turbulence intensity}
\label{BullShit1100}

By using the expansion in powers of $\OBN$ (\ref{ExpansionX}) and Eq. (\ref{ForcingCorrelbis}) and after a long, but straightforward algebra, we can obtain the turbulence intensity in the shear direction as follows:
\EQ
\label{Vxweak}
\langle v_x^2 \rangle = \frac{\tau_f}{(2\pi)^3 \A} \int d^3 k  \phi_{11}({\bf k}) \left[L_0({\bf k}) + \beta^2 \OB L_1({\bf k})\right] \; .
\EN
Here:
\EQA
\label{Bullshit4}
L_0({\bf k}) &=& \int_a^{+\infty} d\tau \frac{e^{-2\xi \left[Q(\tau)-Q(a)\right]}}{(\gamma+\tau^2)^2}  d\tau \; , \\ \nonumber 
L_1({\bf k}) &=& \int_a^{+\infty} d\tau \frac{e^{-2\xi \left[Q(\tau)-Q(a)\right]}}{(\gamma+\tau^2)^2} \left[\tau \{\T(\tau)-\T(a)\} - \frac{1}{2} \ln\left(\frac{\gamma+\tau^2}{\gamma+a^2}\right) \right]  d\tau \; , \\ \nonumber
\T(x) &=& \frac{1}{\sqrt{\gamma}} \arctan\left(\frac{x}{\sqrt{\gamma}}\right) \; .
\ENA
 
In the strong shear limit ($\xi \ll 1$), the integrals $L_0$ and $L_1$ in Eq. (\ref{Bullshit4}) can be simplified:
\EQA
\label{Bullshit5}
L_0({\bf k}) &\sim&  \int_a^{+\infty}  \frac{1}{(\gamma+\tau^2)^2}  d\tau = \frac{1}{2 \gamma} \left[\frac{\pi}{2 \sqrt{\gamma}} - \T(a) - \frac{a}{\gamma+a^2} \right]  \; , \\ \nonumber 
L_1({\bf k}) &\sim& \int_a^{+\infty} \frac{1}{(\gamma+\tau^2)^2} \left[\tau \{\T(\tau)-\T(a)\} - \frac{1}{2} \ln\left(\frac{\gamma+\tau^2}{\gamma+a^2}\right) \right]  d\tau \\ \nonumber
&=& \int_a^{+\infty} \left[\frac{\tau}{2\gamma(\gamma+\tau^2)}+\frac{1}{2 \gamma}\T(\tau)\right] \{\T(\tau)-\T(a)\} d\tau \; . 
\ENA
Note that the second formula for $L_1$ in Eq. (\ref{Bullshit5})  was obtained by integration by part. The leading order behavior of Eq. (\ref{Vxweak}) coming from the term involving $L_0$ is due to shearing effect, showing that $\langle v_x^2 \rangle$ is quenched by flow shear  $ \propto \A^{-1}$ (see \cite{Kim05}). The effect of rotation appears as a correction proportional to $L_1$. One can see from Eq. (\ref{Bullshit5}) that this correction $L_1$ is positive for all values of $a$ (for $a < 0$, the negative part of the integral is always smaller than the positive one as the first term is odd in $\tau$ and the second one is an increasing function of $a$). Therefore, the turbulence intensity $\langle v_x^2 \rangle$ in Eq. (\ref{Vxweak}) increases for $\OB > 0$ whereas it decreases for $\OB < 0$. This can physically be understood from the linear instability analysis (performed in appendix): that is, instability ($\OB>0$) increases turbulence level while stability ($\OB<0$) reduces it.

The other components of the turbulence amplitude can be obtained by following similar analysis in the strong shear limit ($\xi \ll 1$) as follows:
\EQA
\nonumber
\langle v_y^2  \rangle &\sim& \frac{\tau_f}{(2\pi)^3 \A} \int d^3 k  \left[\beta^2 \left(\frac{\pi}{2\sqrt{\gamma}} - \T(a)\right)^2 \phi_{11}({\bf k}) + \phi_{22}({\bf k})\right] \frac{\beta^2}{3\gamma^2} \left(\frac{3}{2\xi}\right)^{1/3} \; , \\ 
\label{Vzweak}
&& \qquad \qquad  \times \left[\Gamma(1/3) + \OB \beta^2 \Gamma(4/3) (-\ln\xi)\right] \;  \\ \nonumber  
\langle v_z^2  \rangle &\sim& \frac{\tau_f}{(2\pi)^3 \A} \int d^3 k  \left[\beta^2 \left(\frac{\pi}{2\sqrt{\gamma}} - \T(a)\right)^2 \phi_{11}({\bf k}) + \phi_{22}({\bf k})\right] \frac{1}{3\gamma^2} \left(\frac{3}{2\xi}\right)^{1/3} \,  \\ \nonumber
&& \qquad \qquad  \times \left[\Gamma(1/3) + \OB \beta^2 \Gamma(4/3) (-\ln\xi)\right] \; . \\ \nonumber  
\ENA
Here, $\Gamma$ is the Gamma function. The first terms in Eq. (\ref{Vzweak}) represent the turbulence amplitude in the direction perpendicular to shear without rotation \cite{Kim05}, which are reduced as $\A ^{-2/3}$ for strong shear. Compared to the leading order behavior of $\langle v_x^2 \rangle \propto A^{-1}$ in shear direction, the reduction is weaker by a factor of $\xi^{1/3}$. That is, a strong anisotropy in turbulence level can be induced for strong shear. The second terms in Eq. (\ref{Vzweak}) capture the effect of weak rotation on sheared turbulence, with turbulence amplitude again being increased or decreased depending on the sign of $\OB$. Furthermore, the correction comes with a multiplying factor $\propto |\ln \xi| >1$, which is larger compared to that for the amplitude in the shear ($x$) direction (which is independent of shear [Eq. (\ref{Vxweak})]). Therefore, in the stable situation ($\OB < 0$) of our interest, weak rotation has the effect of reducing turbulence in the $y-z$ plane more than the one in the shear direction. As a result, the anisotropy induced by flow shear is weakened by rotation. Interestingly, this illustrates the tendency of rotation of leading to almost isotropic turbulence. 

It is also interesting to note that the leading order terms in $\langle v_y^2 \rangle$ and $\langle v_z^2 \rangle$, although apparently very similar, are not exactly the same. For instance, in the case of an isotropic forcing, the angular integration gives $\langle v_y^2 \rangle > \langle v_z^2 \rangle$. This slight anisotropy in $y-z$ (stream and span-wise) directions in sheared turbulence was also observed in numerical simulations of homogeneous turbulence subject to high shear rate: the fluctuating velocity in the direction of the flow is larger than the one in the direction of the shear \cite{Lee90}. This can be contrasted to the exact equipartition between $\langle v_y^2 \rangle$ and $\langle v_z^2 \rangle$ [see (\ref{TurbulenceStrongShear})] in the case of rapid rotation. This is another manifestation of the difference between shear flow and rotation in inducing anisotropic turbulence.

In summary, in the case of a weak rotation/strong shear turbulence ($\A \gg \vert \Omega \vert$ and $\A \gg \nu k_y^2$), the rotation tends to reduce the anisotropy in sheared turbulence.

\subsubsection{Transport of angular momentum}
As noted previously, a strong anisotropy in turbulence is caused by strong shear in the weak rotation limit. There is thus a possibility that this anisotropic turbulence gives rise to non-trivial non-diffusive momentum transport. This will be shown to be the case below.

In the strong shear limit ($\xi \ll  1$), momentum flux can be derived as:
\EQA
\label{LambdaEquator}
\langle v_x v_y  \rangle &\sim& \frac{\tau_f}{(2\pi)^3 \A} \int d^3 k  \Bigl\{ \frac{\phi_{11}({\bf k})}{\gamma} \left[-\frac{1}{2(\gamma+a^2)} + \beta^2 \left(\frac{\pi}{2\sqrt{\gamma}} - \T(a)\right)^2 \right] \\ \nonumber
&& \qquad + \frac{\beta^2 \OB}{3 \gamma} (-\ln\xi)  \left[\beta^2 \left(\frac{\pi}{2\sqrt{\gamma}} - \T(a)\right)^2 \phi_{11}({\bf k}) + \phi_{22}({\bf k})\right] \Bigr\} \; .
\ENA
The momentum flux in (\ref{LambdaEquator}) consists of a diffusive part (the first half term in the integrand on the RHS) and a non-diffusive part (the second half term in the integrand on the RHS). First, the diffusive part, independent of $\OB$, recovers the eddy viscosity of sheared turbulence without rotation \cite{Kim05}, showing that its value decreases as $\propto \A^{-2}$ for strong shear. This result agrees with previous studies of non-rotating sheared turbulence \cite{Nazarenko00b} which found a Reynolds stress inversely proportional to the shear, leading to a log dependence on the distance to the wall for the large-scale shear flow. Second, the non-diffusive part, the correction due to the rotation, is proportional to $\OB$ and is odd in the rotation. This is a non-diffusive contribution to Reynolds stress -- the so-called $\Lambda$-effect. The origin of this non trivial $\Lambda$-effect is the strong anisotropy induced by shear flow on the turbulence even when the driving force is isotropic. It is important to contrast this to the case of rapid rotation limit where non-diffusive fluxes emerge only for anisotropic forcing. A similar result was also found in \S \ref{EqMomentWKB} [see Eqs. (\ref{nutWKBequatorWeak}) and (\ref{nutWKBequator})]. This $\Lambda$-effect [the second term in Eq. (\ref{LambdaEquator})]  is obviously of the same sign as $\OB$ whereas the turbulent viscosity  [the first term in Eq. (\ref{LambdaEquator})] can either be positive or negative, depending on the relative magnitude of the two terms inside the integral. In the two-dimensional (2D) limit with $k_z = 0$ ($\beta=0$), we can easily show that the turbulent viscosity is negative. Note that in this 2D case, $\nu_T<0$ signifies the amplification of shear flow while $|\nu_T| \propto \A^{-2}$ reflects that the generation of shear flow slows down for strong shear. In contrast, in 3D with an isotropic forcing, the turbulent viscosity is positive. Finally, we note that our results here are compatible with previous studies which showed that non-diffusive fluxes of angular momentum \cite{Rudiger80,Kichatinov86b} are proportional to the anisotropy in the background turbulence, which is induced by flow shear in our case. 

\subsubsection{Transport of particles}
Transport of particles has been shown to be severely quenched by shear flow with strong anisotropic properties \citep{Kim05}. We now examine how (weak) rotation affects this.
In the strong shear limit ($\xi \ll 1$), we can find turbulent diffusivity of  particles as:
\EQA
\label{TransportEquatWeak}
D_T^{xx} &\sim& \frac{\tau_f}{(2\pi)^3 \A^2} \int d^3 k  \; \phi_{11}({\bf k}) \left(\frac{\pi}{2\sqrt{\gamma}} - \T(a)\right)^2 \left[1+\OB \beta^2 \,  \frac{-\ln\xi}{3} \right] \; , \\ \nonumber 
D_T^{zz} &\sim& \frac{\tau_f}{(2\pi)^3 \A^2} \int d^3 k  \left[\frac{\phi_{11}({\bf k})\beta^2}{\gamma^2} \left(\frac{\pi}{2\sqrt{\gamma}} - \T(a)\right)^2 + \frac{\phi_{22}({\bf k})}{\gamma^2}\right] \times \\ \nonumber
&& \qquad \frac{1}{3} \left(\frac{3}{2\xi}\right)^{2/3} \Gamma(2/3) \left\{1 + 2 \OB \beta^2 \,  \frac{-\ln\xi}{3} \right\} \; .
\ENA
The first terms in Eq. (\ref{TransportEquatWeak}) manifest the quenching of particle transport for strong shear as $D_T^{xx} \propto \A^{-2}$ and $D_T^{zz}\propto \A^{-4/3}$, with effectively faster transport in span-wise direction compared to shear direction. That is, a strong anisotropic transport can arise for strong shear. It is interesting to contrast this result to that in the case of rotation where the transport in the shear ($x$) direction was larger only by a factor $2$ than the one in the perpendicular direction. The second, correction terms in Eq. (\ref{TransportEquatWeak}) represent the effect of rotation and are proportional to $\OB$: Thus, for $\OB > 0$, the transport is increased whereas it is reduced for $\OB < 0$. This is physically because a weak rotation destabilizes sheared turbulence for $\OB > 0$ whereas it stabilizes for $\OB < 0$ (see figure \ref{Vyhomog} and the discussion in Appendix \ref{LinStab}). Note that a similar behavior was also found in turbulence intensity, given in Eqs. (\ref{Vxweak}) and (\ref{Vzweak}). Thus, one can see that for stable configuration ($\OB <0$) of our interest, the corrections due to rotation tend to weaken the strong anisotropy induced by flow shear. These results highlight the crucial role of shear in transport, in particular in introducing anisotropy. 


To summarize Sec. 3.2, in the slow rotation limit where turbulence is mainly governed by flow shear, turbulence intensity [Eqs. (\ref{Vxweak}) and (\ref{Vzweak})] and transport [Eq. (\ref{TransportEquatWeak})] can be severely quenched with strong anisotropy due to shearing while weak rotation weakens this anisotropy to next order. The strong anisotropic turbulence was shown to give rise to a $\Lambda$-effect for momentum transport [Eq. (\ref{LambdaEquator})] even for an isotropic forcing.

\section{Discussion}
\label{Implications}
In \S \ref{Equator}, depending on the values of the parameter $\xi=(\nu k_y^2)/\A$, we considered two regimes: the strong shear ($\xi \ll 1$) and the weak shear limits ($\xi \gg 1$). Since we are interested in the effects of flow shear as well as rotation, we here summarize and discuss our results obtained in the limit of strong shear with $\xi \ll 1$. Table \ref{Summary} summarizes our findings by highlighting the quenching of these quantities due to large shearing rate $\A$ and the rotation rate $\Omega$ (or their ratio, $\OB=\Omega / \A$). These results are discussed in the following.

\renewcommand{\arraystretch}{1.8}
\begin{table}
\begin{center}
\begin{tabular}{|l|c|c|}
\hline
& \;  $\Omega \gg \A \; $ & $ \; \Omega \ll \A \; $ \\ \hline
$ \; \langle v_x^2 \rangle \;$ & $ \qquad \A^{-1} \qquad $ & $\A^{-1} \left[1+ C \OB\right]$ \\
$ \; \langle v_y^2 \rangle \sim \langle v_z^2 \rangle \; $ & $ \A^{-1} \vert \ln \xi \vert$  & $\A^{-2/3} \left[1+ C  \OB \vert \ln \xi \vert\right]$ \\ \hline
$  \; \nu_T  \; $ & $-\A^{-2}$ & $\A^{-2} $ \\ \
$  \; \Lambda_x  \; $ & $0$ & $\A^{-2} \vert \ln \xi \vert$ \\ \hline
$ \; D_T^{xx} \; $ & $\Omega^{-1}$ &  $\A^{-2} \left[1+ C \OB \vert \ln \xi \vert\right]$ \\
$ \; D_T^{yy} \sim D_T^{zz} \; $ & $\Omega^{-1}$ & $\A^{-4/3} \left[1+ C \OB \vert \ln \xi \vert\right]$ \\  \hline
\end{tabular}
\end{center}
\caption{\label{Summary} Summary of our results obtained in the strong shear limit ($\xi = \nu k_y^2/\A \ll 1$). The $C$ symbol stands for an additional constant of order $1$.}
\end{table}

\subsection{Turbulence amplitude}
In all the cases considered, turbulence amplitude is always quenched due to strong shear ($\xi =\nu k_y^2 / \A \ll 1$), with stronger reduction in the direction of the shear ($x$) than those in the perpendicular directions. Specifically, in the large rotation limit, they scale as $\A ^{-1}$ and $\A^{-1} \vert \ln \xi \vert$, respectively while in the weak rotation limit, they scale as  $\A^{-1}$ and $\A^{-2/3}$, respectively. Thus, flow shear always leads to weak turbulence with an effectively stronger turbulence in the plane ($y$-$z$) than in the shear direction, regardless of rotation rate. The anisotropic reduction of turbulence amplitude is because of the shear which increases the dissipation (anisotropically) by efficiently creating small-scale fluctuations in the $x$-direction, with a direct impact on turbulence in the shear direction (see Figure \ref{ShearEff}). The anisotropy in turbulence amplitude is however weaker by a factor of $\xi^{1/3} \vert \ln \xi \vert$ ($\propto \A^{-1/3} \vert \ln \xi \vert$) in the rapid rotation limit than that in weak rotation limit since rotation favors almost-isotropic turbulence. In the case of weak rotation, the effect of shear on turbulence amplitude can be understood in terms of stability of rotating shear flow (see Appendix \ref{LinStab} for more details). In the case of weak rotation ($\Omega \ll \A$), the effect of rotation appears in combination with the linear instability criterion in turbulence amplitude with linear stability $\OB<0$ (instability $\OB>0$) decreasing (increasing) turbulence amplitude. For stable configuration $\OB<0$, the rotation thus has the effect of weakening the anisotropy caused by strong shear. In summary, turbulence amplitude is quenched by shear with strong anisotropy while rotation tends to weaken the shear-induced anisotropy.

\subsection{Transport of angular momentum}
The transport of angular momentum was found to involve two contributions: the turbulent viscosity $\nu_T$ and the $\Lambda$-effect. The former is a diffusive flux, making the effective viscosity to $\nu_T + \nu$ ($\nu$ is the molecular viscosity) while the latter is a non-diffusive momentum flux. The turbulent viscosity is negative with inverse cascade of energy as long as rotation is sufficiently strong compared to flow shear ($\Omega \gg A$). This is consistent with previous works which showed that a turbulent viscosity exists only for highly anisotropic flows \cite{Dubrulle91}  or two-dimensional flows \cite{Gama94}. As rotation tends to make flow two-dimensional, we expect the turbulent viscosity to be negative. In comparison, turbulent viscosity is positive in the opposite limit of weak rotation ($\Omega \ll \A$). This is because rotation favors transfer of energy from small scales to large scales (inverse cascade) while flow shear efficiently creates small scales via shearing, cascading the energy from large to small scales. Even if the eddy viscosity is negative for strong rotation ($\Omega \gg \A$), flow shear, which transfers energy to small scales, has an interesting effect by slowing down the rate of inverse cascade with the value of negative eddy viscosity decreasing as $|\nu_T| \propto \A^{-2}$ for strong shear.

The non-diffusive part of momentum transport ($\Lambda$-effect) can act as a source of large-scale flow, preventing a uniform rotation to be solution of the averaged Reynolds equation. A strong anisotropy induced by flow shear \cite{Kim05} gives rise to non-trivial $\Lambda$-effect even for an isotropic forcing. Note that in the absence of flow shear, the appearance of a $\Lambda$-effect requires a source of anisotropy in the system such as an anisotropic forcing in which case the $\Lambda$-effect is proportional to the anisotropy in the velocity field \cite{Kichatinov86b,Rudiger89}. Interestingly, our results show that the $\Lambda$-effect scales as $\A^{-2} \vert \ln \xi \vert$ whereas the anisotropy in the velocity amplitude is given, to leading order, by $\A^{-4/3}$. Consequently, the $\Lambda$-effect is smaller than the anisotropy in the turbulent velocity amplitude. This is because the anisotropy is not simply given here but has to be induced self-consistently by the shear during the evolution. In other words, the anisotropy does not remain the same at all time, and the resulting $\Lambda$-effect is smaller than the anisotropy in the velocity amplitude in the long-time limit. One can also note that the magnitude of the $\Lambda$-effect is not the same in the two cases. 
 
\subsection{Transport of particles}
The dynamics of particles transport crucially depends on whether rotation is stronger or weaker than flow shear. When rotation is stronger than flow shear ($\Omega \gg \A$), the transport is inhibited by inertial waves, being quenched inversely proportional to the rotation rate (i.e. $\propto \Omega^{-1}$) while in the opposite case where flow shear is stronger than rotation, it is reduced by shearing as $\A^{-1}$. It  is important to compare this result with turbulence amplitude, which is quenched by shearing even when $\Omega \gg \A$. This strikingly different behavior between particle transport and turbulence amplitude highlights the different roles of waves and flow shear in turbulence regulation; that is, waves mainly affect transport by altering phase relation while flow shear quenches both transport and turbulence level, via enhanced dissipation.
   
Furthermore, in the strong rotation limit ($\Omega \gg \A$) where the transport of particles is dominated by inertial waves, the transport is almost isotropic with only a slight anisotropy -- the transport in the direction parallel to the rotation is twice larger than the one in the perpendicular direction (see also Eq. (\ref{Isotropic2}) and \cite{Kichatinov94}). However, in the weak rotation limit, it is flow shear that quenches particle mixing; the anisotropy in resulting transport can be very large with much slower mixing by a factor of $A^{-2/3}$ in the direction of shear. The rotation on shear-dominated turbulence weakens the anisotropy. 

\subsection{Effect of a bounded domain}
\label{BoundedDomain}
The calculation of all the turbulent coefficients in the weak shear limit ($\xi \gg 1$) and also of the transport of particles in the strong shear limit ($\xi \ll 1$) required the evaluation of the integrals of the following type:
\EQ
\label{Bullshit300}
I({\bf k},\Omega) = \int \frac{H(k)}{\nu^2 k^4 + \overline{\omega_0}^2} \, d^3 k \; ,
\EN
where $\overline{\omega_0} = ({\bf \Omega \cdot k}) / k$ is the projection of the unit vector in the direction of the wave number on the rotation axis. When the domain of integration is unbounded (infinite), the integration over the angular variable of this integral becomes proportional to $\Omega^{-1}$, when the rotation rate $\Omega$ is sufficiently large [see Eq. (\ref{Bullshit200})-(\ref{nutWKBequatorWeak}) for details]. This is because this integral involves some contribution of order unity (when ${\bf \Omega \cdot k} = 0$) and others of magnitude $\Omega^{-2}$.

However, in realistic situations, the domain of integration in Fourier space is bounded with a minimum wavenumber that is permitted in the system (corresponding to a maximum length, for instance the size of the box) in the direction of the rotation. If we call this minimum wavenumber $k_m = \text{min}(k_x)$, we can show that the preceding scaling of $\Omega^{-1}$ is valid only when $\nu^2 k^6 \gg \Omega^2 k_m^2$. In the opposite case, the term  $\overline{\omega_0}^2$ in Eq. (\ref{Bullshit300}) is always dominant, altering this integral to $\propto \Omega^{-2}$ for large rotation rate, with a stronger dependence on $\Omega$.



\section{Conclusion}
\label{Conclusion}
In this paper, we have performed a thorough investigation of the combined effects of shear and rotation on the structure of turbulence, by using a quasi-linear theory. We assumed an external forcing in the Navier-Stokes equation which leads to an equilibrium situation where the dissipation (whose effect is enhanced by the shear) is balanced by the injection of energy due to forcing. It is useful to recall that there are three (inverse) time-scales in the problem: the shearing rate $\A$, the rotation rate $\Omega$ and the diffusion rate $\mathcal{D}=\nu k_y^2$ where $\nu$ is the (molecular) viscosity of the fluid and $k_y^{-1}$ is a characteristic small-scale of the forcing. The first regime of strong rotation ($\Omega \gg \A$) has been studied in the strong shear ($\A \gg \mathcal{D}$) and weak shear ($\A \ll \mathcal{D}$) limits. However, the second regime of weak rotation has been considered only in the strong shear ($\A \gg \mathcal{D}$) case, as the effects of both shear and rotation disappear in the opposite case. 

While both rotation and (stable) shear flow tend to regulate turbulence, there are important differences in their effects, which should be emphasized. Rotation, by exciting inertial waves, tends to reduce turbulence transport more heavily than turbulence amplitude while shear flows reduce both of them to a similar degree. That is, rotation (or waves) quenches the cross-phase (normalized flux) more than shear flow does \cite{Kim03b,Kim06}. Furthermore, in sharp contrast to rotation, shear flow induces a strong anisotropic turbulence and transport (e.g. momentum transport, chemical mixing, etc.). 

Specifically, in the large rotation limit ($\vert \Omega \vert \gg \A$), we have found:
\begin{itemize}
\item The turbulent intensity is reduced only by a strong shear  (i.e. in the case of strong rotation and strong shear) and in an anisotropic way. 
\item As the dissipation decreases (compared to the shear), there is a crossover from a positive to a negative viscosity. 
\item The transport of particle is reduced by rotation, with a slight anisotropy of a factor $2$, largely unaffected by shear.
\end{itemize}

In the opposite weak rotation limit ($\vert \Omega \vert \ll \A$), we found that the main reduction is due to the shear with an anisotropic turbulence with preferred motion and transport in the plane perpendicular to the shear. Rotation can increase or decrease slightly the turbulence intensity and the particle transport, depending on the sign of $\OB = \Omega / \A$.

Furthermore, we found non-diffusive flux for momentum transport (the so-called $\Lambda$-effect) which transfers energy from the fluctuating velocity field to the large-scale flow. In the large rotation limit, this term can appear only for an anisotropic forcing. In contrast, in the weak rotation limit, rotation acting together with shear flow was shown to give rise to non diffusive fluxes even with an isotropic forcing.

These results can have significant implications for astrophysical and geophysical systems. For instance, the $\Lambda$-effect and/or negative viscosity can provide a mechanism for the generation of ubiquitous large-scale shear flows, which are often observed in these objects. Furthermore, the anisotropic mixing of scalars should be taken into account in understanding the surface depletion of light elements in stars \cite{Pinsonneault97}. Finally, we note that numerical confirmation of our prediction and the extension of our work to stratified rotating sheared turbulence with/without magnetic fields remain challenging important problems, and will be addressed in future publications.

\begin{acknowledgments}
We thank A. P. Newton for providing us with Figure \ref{ShearEff} and L. L. Kichatinov for useful comments. This work was supported by U.K. PPARC Grant No. PP/B501512/1.
\end{acknowledgments}

\begin{appendix}
\section{Linear stability analysis of the homogeneous system}
\label{LinStab}
As Eq. (\ref{quasi-linear}) is the same as that for a perturbation ${\bf u}$ about a basic flow ${\bf U}_0$, up to the extra forcing term ${\bf f}$, our study gives some insight into the stability of shear flows in presence of rotation. After summarizing results previously obtained by others, we present our results in the cases where the rotation and the shear are perpendicular and parallel, respectively.

The case of the plane shear flow in a rotating frame has been studied by many authors focusing on the stability both in the laminar and the turbulent cases. In the case of a rotation vector $\tilde{\bf \Omega} = \tilde{\Omega} {\bf e}_z$  perpendicular to the plane of the shear flow, \cite{Bradshaw69} proposed an analogy between rotation and stratification [supported by calculation of \cite{Pedley69}] and showed that the system was unstable if the vorticity of the shear flow $- \A {\bf e}_z$ is anti-parallel to the rotation and sufficiently strong. Precisely, the ratio $\OB = 2 \tilde{\Omega} / \A$ must lie in the interval $[0 \, , \,  1]$ for instability. This destabilization of laminar shear flow by rotation has a counterpart for turbulent flows where the rotation can stabilize turbulence (by decreasing its kinetic energy) or destabilize it, as shown by \cite{Tritton92} using a displacement argument. It is interesting to note that both Bradshaw and Tritton arguments are pressure-less. However, the Pedley criterion was shown to hold by \cite{Yanase93}, using stability analysis confirmed by simulations \cite{Metais95}: the cyclonic shear ($\OB < 0$) is always stabilizing whereas the anticyclonic shear ($\OB > 0$) is destabilizing for weak rotation while stabilizing for high rotation, in agreement with Bradshaw criterion. These conclusions are confirmed for a Poiseuille flow, both experimentally \cite{Johnston72} and numerically \cite{Kristoffersen93}, and  for a plane Couette flow \cite{Bech96,Bech97}. The fact that the pressure-less argument gives the exact stability criterion is due to the fact that the modes which are dominant in the instability process are naturally unaffected by pressure fluctuations \cite{Leblanc97}.

\begin{figure}
\begin{center}
\includegraphics[scale=0.42,clip]{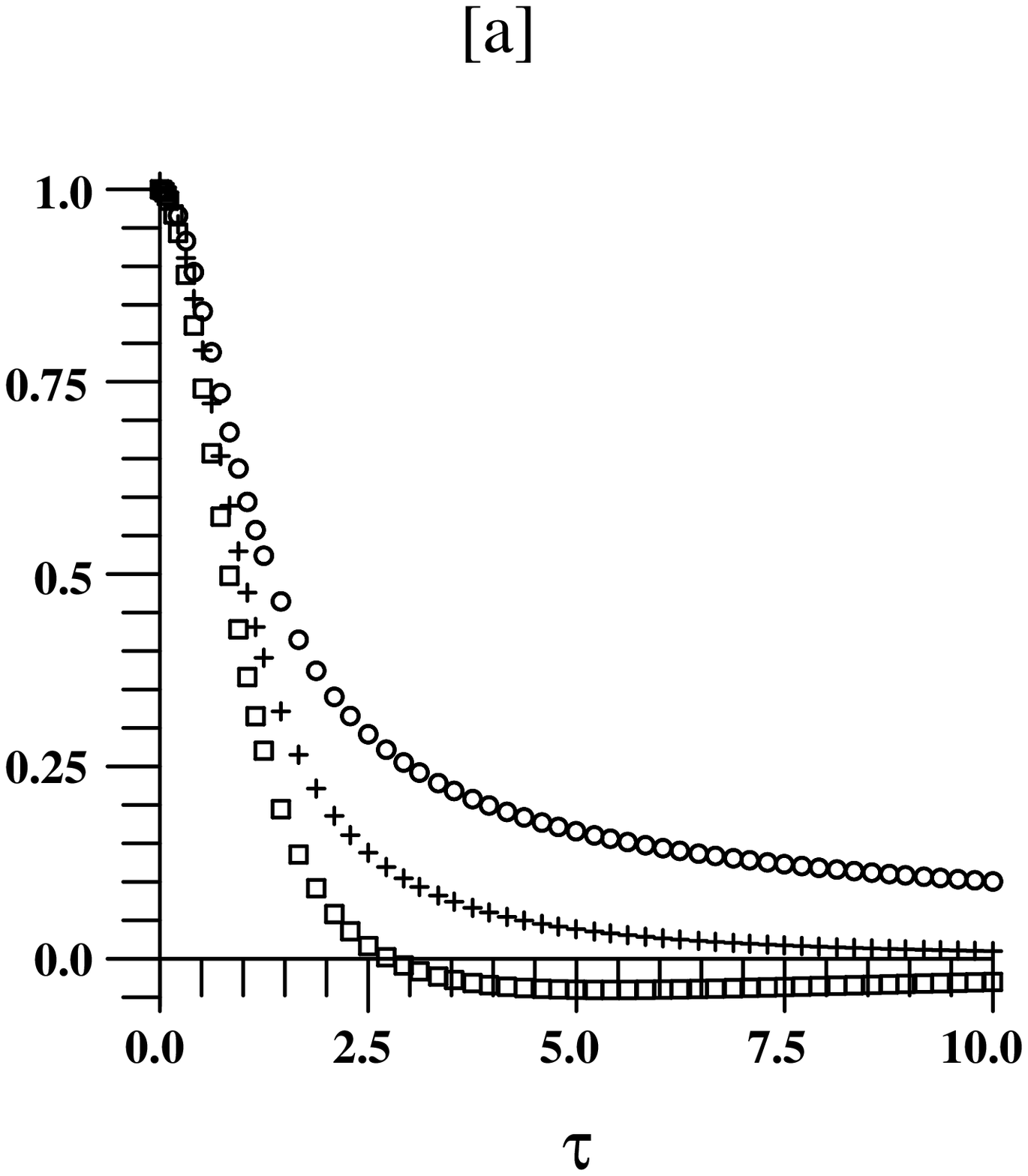}
\includegraphics[scale=0.42,clip]{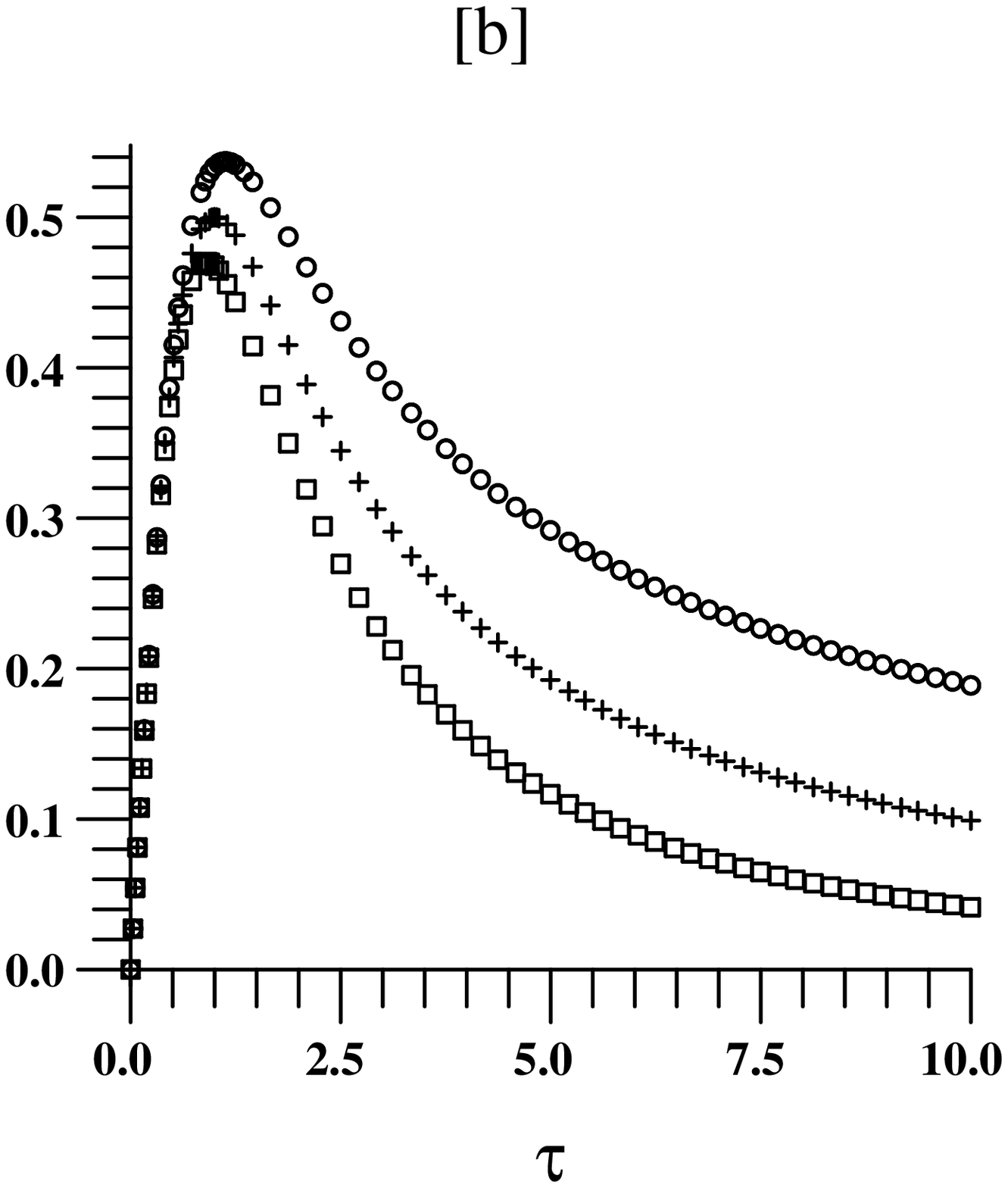}
\caption{\label{Vxhomog} Evolution of the solution $X_1$ (panel [a]) and $X_2$ (panel [b]) as a function of $\tau$ for $b=-0.5$ (circles), $b=0$ (crosses) and $b=0.5$ (squares).}
\end{center}
\end{figure}

The exact solution of the homogeneous part of Eq. (\ref{System3}) for the velocity $\hat{v}_x$ can be found in terms of generalized hyper-geometric function $F([a_1,a_2,\dots],[b_1,b_2,\dots],x)$ \cite{Gradshtein}. Two independent solutions are:
\EQA
\label{Bullshit100}
X_1(\tau) &=& F\Bigl(\Bigl[\frac{3}{4}+\frac{\sqrt{1-4b}}{4},\frac{3}{4}-\frac{\sqrt{1-4b}}{4}\Bigr],\Bigl[\frac{1}{2}\Bigr],-\frac{\tau^2}{\gamma} \Bigr) \; , \\ \nonumber 
X_2(\tau) &=& \tau F\Bigl(\Bigl[\frac{5}{4}+\frac{\sqrt{1-4b}}{4},\frac{5}{4}-\frac{\sqrt{1-4b}}{4}\Bigr],\Bigl[\frac{3}{2}\Bigr],-\frac{\tau^2}{\gamma}\Bigr) \; .
\ENA
Here, $b=\beta^2 \OB (\OB-1)$ is (up to the multiplicative constant $\beta^2$) the quantity introduced by \cite{Bradshaw69} (see discussion in the introduction). Figure \ref{Vxhomog} shows the evolution of these two functions as a function of $\tau$. 

\begin{figure}
\begin{center}
\includegraphics[scale=0.55,clip]{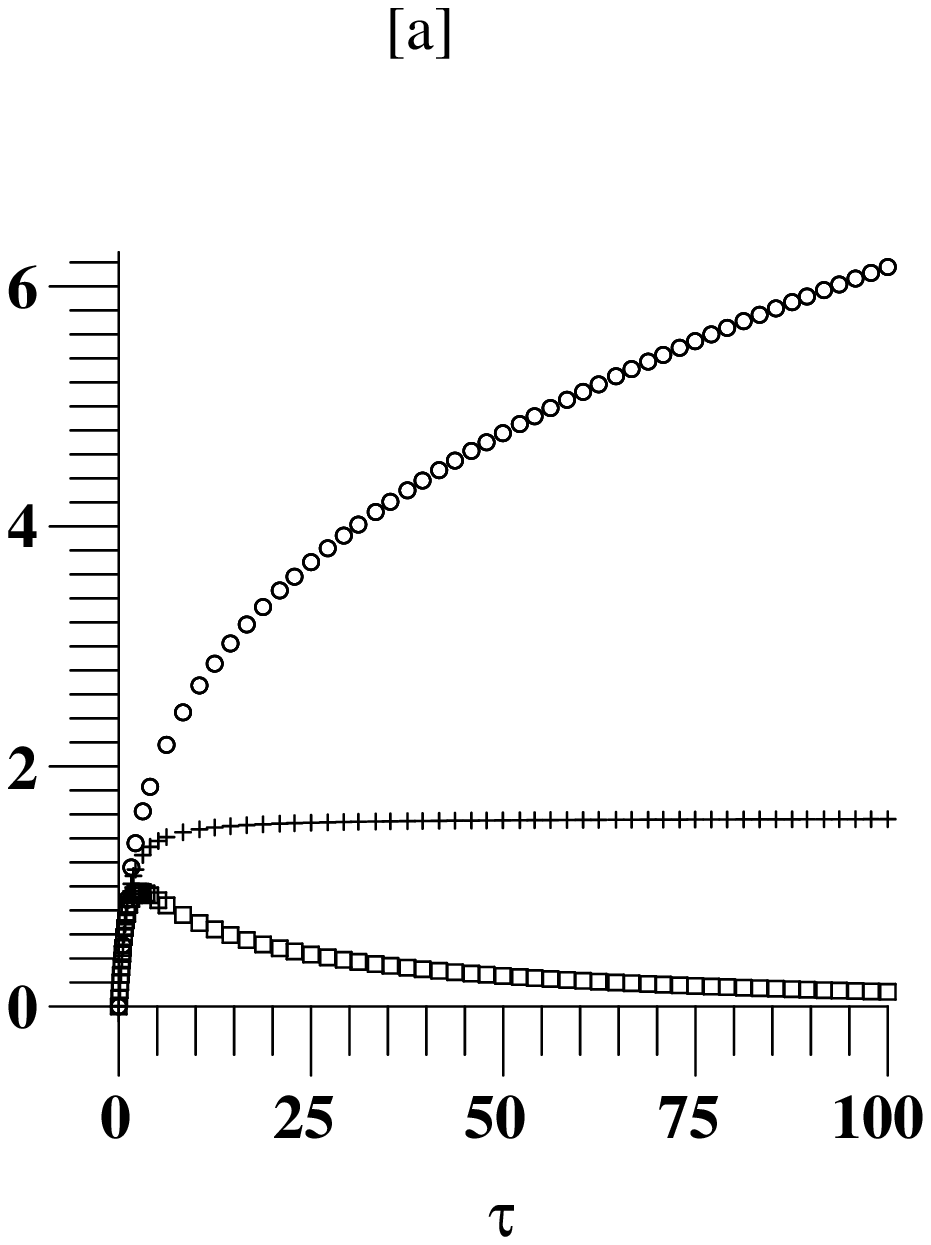}
\hskip0.5cm
\includegraphics[scale=0.55,clip]{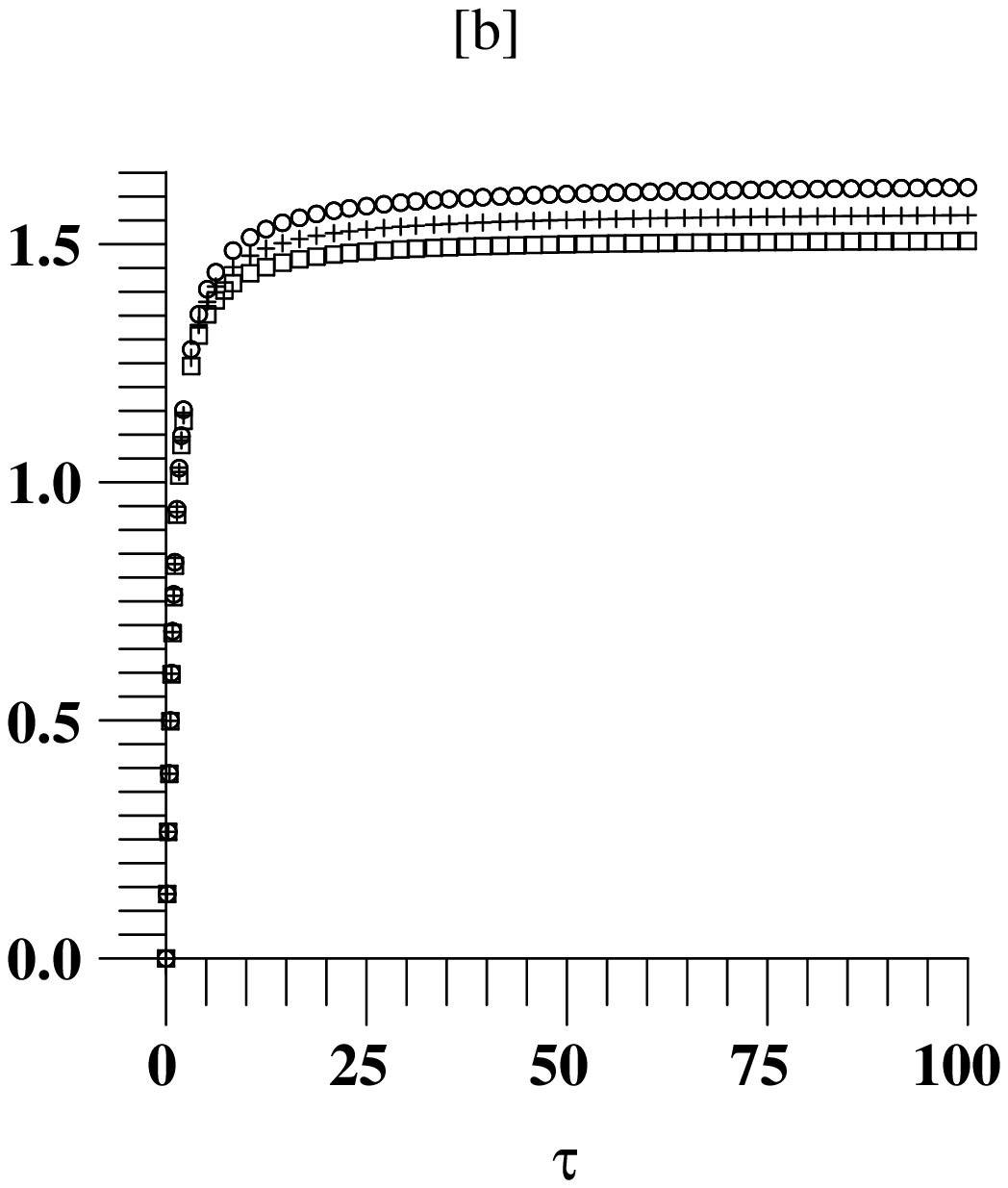}
\caption{\label{Vyhomog} Evolution of the solution $Y_1$ (panel [a]) and $Y_2$ (panel [b]) as a function of $\tau$ for $b=-0.5$ (circles), $b=0$ (crosses) and $b=0.5$ (squares).}
\end{center}
\end{figure}

Solutions for the other components of the velocity are obtained by using the last two equations of (\ref{System3}) :
\EQA
\label{Bullshit101}
\hat{v}_y &=& - \frac{1}{\gamma} \left[\tau  X_n(\tau) + \beta^2 (\OB-1) Y_n(\tau) \right] \; , \\ \nonumber
\hat{v}_z &=& - \frac{\beta}{\gamma} \left[\tau  X_n(\tau) - (\OB-1) Y_n(\tau) \right] \; ,
\ENA 
for $n=1$ or $2$. Here, $Y_1$ and $Y_2$ are defined as:
\EQA
Y_1(\tau) &=& \tau F\Bigl(\Bigl[\frac{3}{4}+\frac{\sqrt{1-4b}}{4},\frac{3}{4}-\frac{\sqrt{1-4b}}{4}\Bigr],\Bigl[\frac{3}{2}\Bigr],-\frac{\tau^2}{\gamma} \Bigr) \; , \\ \nonumber 
Y_2(\tau) &=&  -\frac{\gamma}{b} F\Bigl([\frac{1}{4}-\frac{\sqrt{1-4b}}{4}, \frac{1}{4}+\frac{\sqrt{1-4b}}{4}], [\frac{1}{2}], -\frac{\tau^2}{\gamma} \Bigr) \; .
\ENA
The plots of $Y_1(\tau)$ and $Y_2(\tau)$ are shown in figure \ref{Vyhomog}.

Figure \ref{Vyhomog} shows that the eigenfunctions diverge for $\tau \rightarrow \infty$ when $b < 0$. This is because  shear flows in presence of rotation (perpendicular to the shear flow) is stable only for $b > 0$. This result agrees with \cite{Bradshaw69} and \cite{Salhi97}. We can also notice that the solution with $b > 0$ always decays faster than that with $b < 0$.

In conclusion, we recovered the Bradshaw criterion \cite{Bradshaw69}. In our notation, it states that the configuration is unstable if $B= -\OB(1-\OB) < 0$ or, equivalently, if $\OB = \Omega/ \A$ lies in the interval $[0 \, , \, 1]$. This result has already been reported by many authors, who showed not only that the maximum destabilization occurs for $\OB = 1/2$ but also that there is an important asymmetry with respect to $\OB = 1/2$ which is not included in the Bradshaw criterion \cite{Speziale89,Cambon94,Salhi97}. This is because Bradshaw criterion can be recovered by a pressure-less analysis: while the pressure does not affect the most unstable modes \cite{Leblanc97} and thus does not alter the instability criterion, is does destroy this symmetry. We can easily show that there is indeed asymmetry with respect to $\OB = 1/2$ in our results: even if Eq. (\ref{Bullshit100}) for the $x$-component of the velocity is symmetric with respect to $\OB = 1/2$ (as it depends only on $b = -\beta^2 B$), Eq. (\ref{Bullshit101}) for the other components of the velocity are not because of the term proportional to $\OB - 1$. 

\section{Derivation of the particle transport in the large rotation limit}
\label{Appendix2}
By using Eqs. (\ref{EquTransport}), (\ref{WKBEquator}) and (\ref{ForcingCorrelbis}), we obtain the transport of particles in the direction of the shear as follows:
\EQA
\label{ChemicalsLargeRotation}
D_T^{xx} &=& - \frac{\tau_f}{(2 \pi)^3 \A^2} \int d^3 k \;  \gamma (\gamma + a^2)^{1/4} F(k) \; \frac{S^0_3}{\omega_0} \; , \\ \nonumber
D_T^{yy} &=& \frac{\tau_f}{(2 \pi)^3 \A^2} \int d^3 k \;  \frac{\sqrt{\gamma + a^2} F(k)}{\gamma \omega_0} \; \times  \\ \nonumber
&& \quad \Bigl\{ - \frac{a S_3^1}{(\gamma+a^2)^{1/4}} + \frac{\beta \theta}  {(\gamma+a^2)^{1/4}} \left (\sqrt{\gamma+a^2} C_3^1 - a C_1^0 \right) - \beta^2 (\gamma+a^2)^{1/4} S_1^0 \\ \nonumber 
&&  \quad + \frac{\beta \theta}{2(\gamma+a^2)^{3/4}} \left(\beta^2 \sqrt{\gamma+a^2} \mathcal{C}_1^0 - a \mathcal{C}_3^1 \right) + \frac{\beta^2}{2(\gamma+a^2)^{3/4}} \left(a \mathcal{S}_1^0 + \sqrt{\gamma+a^2} \mathcal{S}_3^1 \right) \Bigr\} \; , \\ \nonumber
\ENA
where:
\EQA
\label{IntegralsParticles}
\zeta_n^p({\bf k}) &=& \int_a^{+\infty}  \frac{\tau^p e^{-2\xi \left[Q(\tau)-Q(a)\right]}}{(\gamma+\tau^2)^{n/4}}  \exp\left[i \omega_0 \left\{ s(a)-s(\tau) \right\} \right] d\tau \; , \\ \nonumber
\mu_n^p({\bf k}) &=& \int_a^{+\infty}  \frac{\tau^p e^{-2\xi \left[Q(\tau)-Q(a)\right]}}{(\gamma+\tau^2)^{n/4}} (\tau-a) \exp\left[i \omega_0 \left\{ s(a)-s(\tau) \right\} \right] d\tau \; , \\ \nonumber
C_n^p &=& \Re(\zeta_n^p) \; , \quad S_n^p = \Im(\zeta_n^p) \; , \quad \mathcal{C}_n^p = \Re(\mu_n^p) \; , \quad \mathcal{S}_n^p = \Im(\mu_n^p) \; . 
\ENA
The expression for $D_T^{zz}$ is omitted here as it is very similar to that for $D_T^{yy}$. The asymptotic behavior of integrals (\ref{IntegralsParticles}) can be obtained to leading order in $\OBN^{-1}$ as:
\EQ
\label{IntegralsParticlesAsympt}
\zeta_n^p({\bf k}) \sim  \frac{a^p (2 \nu k^2 - i \overline{\omega_0}) \A}{(\gamma+a^2)^{n/4} [4\nu^2 k^4 + \overline{\omega_0}^2]} \; , 
\EN
where $\overline{\omega_0} = \omega_0 \A / \sqrt{\gamma+a^2}$. In comparison, the functions $\mu_n^p$ vanish to leading order and are thus omitted here. By using Eq. (\ref{IntegralsParticlesAsympt}) in Eq. (\ref{ChemicalsLargeRotation}), we obtain the following results:
\EQA
\label{ChemicalsLargeRotation2}
D_T^{xx} &=& \frac{\tau_f}{(2 \pi)^3} \int d^3 k \;   F(k) \frac{\gamma}{\gamma+a^2} \frac{1}{4\nu^2 k^4 + \overline{\omega_0}^2} \; , \\ \nonumber
D_T^{yy} &=& \frac{\tau_f}{(2 \pi)^3} \int d^3 k \;  F(k) \frac{a^2+\beta^2}{(\gamma+a^2)} \frac{1}{4\nu^2 k^4 + \overline{\omega_0}^2} \; , \\ \nonumber
D_T^{zz} &=& \frac{\tau_f}{(2 \pi)^3} \int d^3 k \;   F(k) \frac{1+a^2}{(\gamma+a^2)} \frac{1}{4\nu^2 k^4 + \overline{\omega_0}^2} \; .
\ENA
Here, we have discarded all the terms which are odd in $a$ (for example in $D_T^{yy}$, the terms proportional to $C_1$ and $C_3$) as they vanish after angular integration. After performing this integration, (\ref{ChemicalsLargeRotation2}) reduces  to Eq. (\ref{Isotropic2}) given in the main text. 

\end{appendix}

\bibliographystyle{apsrev}
\bibliography{../../../Biblio/Bib_sun,../../../Biblio/Bib_shear,../../../Biblio/Bib_Geo,../../../Biblio/Bib_turbu,../../../Biblio/Bib_maths,../../../Biblio/Bib_dynamo,Bib_RotShear}

\end{document}